\documentclass[12pt]{article}
\usepackage{epsfig}
\begin{document}
\setlength{\oddsidemargin}{0in}
\setlength{\textwidth}{17cm}
\setlength{\topmargin}{-0.5in}
\setlength{\textheight}{22cm}
\newcommand{\be}{\begin{equation}}
\newcommand{\ee}{\end{equation}}
\newcommand{\bea}{\begin{eqnarray}}
\newcommand{\eea}{\end{eqnarray}}
\newcommand{\rotatefigure}[2]
{
\begin{figure}[t]
  \begin{center}
  \mbox{\begin{turn}{-90}\psfig{file=#1,height=60 ex}
  \end{turn}}
  \end{center}
  \caption{{\small #2}}
\end{figure}
}
\setlength{\unitlength}{.6 ex}
\newcommand{\vertice}{\mbox{ 
\begin{picture}(14,6)
	 \put(8,2){\line(1,1){20}}
 \put(8,2){\line(1,0){39}}
 \put(8,2){\line(1,-1){20}}
 \put(47.6,2){\line(-1,1){20}}
 \put(47.6,2){\line(-1,-1){20}}
 \put(27.8,21.8){\line(0,1){10}}
 \put(27.8,-17.8){\line(0,-1){10}}
 \put(15.,14.){\makebox(0,0){$ p $}}
 \put(48.,14.){\makebox(0,0){$1-p$}}
 \put(29.5,-0.8){\makebox(0,0){$p-k$}}
 \put(14.5,-10.){\makebox(0,0){$ k $}}
 \put(48.,-10.){\makebox(0,0){$1-k $}}
 \put(24.,29.8){\makebox(0,0){$ 1 $}}
 \put(24.,-25.8){\makebox(0,0){$ 1 $}}
\end{picture}}}
\title{\bf Interaction of Reggeized Gluons in the
Baxter-Sklyanin Representation}
\author{{\bf H. J. de Vega$^{(a)}$,   L. N. Lipatov$^{(b,\diamond)}$ } \\ \\
\small
(a)LPTHE\footnote{Laboratoire Associ\'{e} au CNRS UMR 7589.},
\small Universit\'e Pierre et Marie Curie (Paris VI) \\
\small et Denis Diderot  (Paris VII),  Tour 16, 1er. \'etage, \\ \small 4,
Place Jussieu 75252 Paris, Cedex 05, France \\
\small (b)Petersburg Nuclear Physics Institute, \\
\small Gatchina, 188300, St.Petersburg, Russia and \\
\small  LPMT Universit\'e Montpellier 2, \\
\small Place Eug\`ene Bataillon, 34095 Montpellier Cedex 05, France}
\date{\today}
\maketitle
\begin{abstract}
We investigate the Baxter equation for the Heisenberg spin model
corresponding to a generalized BFKL equation describing
composite states of $n$ Reggeized gluons in the multi-color limit of
QCD.
The Sklyanin approach is used to find an unitary transformation from the
impact parameter representation to the representation in which the wave
function factorizes as a product of Baxter functions and a
pseudo-vacuum state. We
show that the solution of the Baxter equation is a meromorphic function
with poles $(\lambda -i\,r)^{-(n-1)}$ ($r=0,1,...$) and that the
intercept for the composite Reggeon states  is
expressed through the behavior of the Baxter function around the pole
at $\lambda =i $. The absence of pole singularities in the two
dimensional $\overrightarrow{\lambda}$-plane for the bilinear
combination of holomorphic and anti-holomorphic Baxter functions leads to
the quantization of the integrals of motion because the holomorphic
energy should be the same for all independent Baxter functions.
\end{abstract}
\noindent
$^{(\diamond )}$ {\it Work supported partly
by INTAS grants  1997-31696, 2000-366,  CRDF grant RP1-2108, by NATO
and by the Russian Fund of Fundamental Investigations}
\section{Introduction}

In the leading logarithmic approximation (LLA) of perturbative QCD the
Reggeons (reggeized gluons) move in the two-dimensional
impact parameter plane $\overrightarrow{\rho}$ and interact pairwise
\cite{BFKL,norm}. To unitarize the QCD scattering
amplitudes at high energies one should take into account the multi-Reggeon
exchanges in the $t$-channel. The composite states of the reggeized
gluons satisfy a Schr\"odinger-like equation \cite{BKP}.

The Reggeon Hamiltonian in the infinite color limit $N_c \rightarrow
\infty$
takes a simple form and can be written as
follows \cite{lip1},
\begin{equation}
H=\frac{1}{2}(h+h^{*}) \quad , \quad \left[ h,h^{*}\right] =0\,,
\end{equation}
where the holomorphic and anti-holomorphic Hamiltonians
\begin{equation}
h=\sum_{k=1}^{n}h_{k,k+1\,} \quad , \quad
h^{*}=\sum_{k=1}^{n}h_{k,k+1\,}^{*} \; ,
\end{equation}
are expressed in terms of the pair BFKL operator \cite{BFKL,lip1}:

\begin{equation}
h_{k,k+1}=\log p_k+\log p_{k+1}+\frac 1{p_k}\left( \log \rho _{k,k+1}\right)
p_k+\frac 1{p_{k+1}}\left( \log \rho _{k,k+1}\right) p_{k+1}+2\,\gamma \,.
\label{hyh*}
\end{equation}
Here $\rho _{k,k+1}=\rho _k-\rho _{k+1}\,,\,p_k=i\,{\frac \partial {\partial
\rho _k}}\;,\,p_k^{*}=i\,{\frac \partial {\partial \rho _k^{*}}}\,$, and $%
\gamma =-\psi (1)$ is the Euler-Mascheroni constant.

In this context the Pomeron is a compound state of two reggeized
gluons and the Odderon is constructed from three reggeized gluons.

The operator $h$ is
invariant under the M\"obius transformations \cite{norm} with
generators:
\[
\vec M=\sum_{k=1}^n\vec {M_k}\,;\,\,M_k^3=\rho _k\,\partial
_k\,,\,\,M_k^{-}=\partial _k\,,\,\,M_k^{+}=-\rho _k^2\;\partial _k\;.
\]
The Casimir operator of this group is
\begin{equation}
\vec M^2=-\sum_{l<r}^n\rho _{lr}^2\;
\partial _l\partial _r\,.  \label{casi}
\end{equation}

The Hamiltonian $h$ describes the integrable $XXX$ spin model with the spins
being the generators $\vec M_k$ of the M\"obius group \cite{lip2}. The
integrals
of motion of this model are generated by the transfer matrix which is the
trace of the monodromy matrix satisfying the Yang-Baxter equation \cite
{lip2}. Therefore, the Quantum Inverse Scattering Method \cite{QISM,rev}
can be  applied to find an algebraic solution of the Schr\"odinger
equation.

The pair hamiltonian ({\ref{hyh*}) can be obtained from the fundamental
monodromy matrix associated to the XXX Heisenberg spin model
\cite{lip2,lip3,fk}.
Notice that the local operators $p_k, \rho _{k} $ act in an infinite
dimensional Hilbert space whereas the spin operators in the usual Heisenberg
model are finite dimensional matrices both for integer or half
plus an
integer spin.

The auxiliary $L$-operator for the Heisenberg spin model with $s=-1$ is
given below \cite{lip3,fk,kor}
\begin{equation}
L_k(u)=\left(
\begin{array}{cc}
u+p_k \; \rho_{k0} & -p_k \\
p_k \; \rho _{k0}^2 & u-p_{k} \; \rho_{k0}
\end{array}
\right) \,.
\end{equation}
where $\rho _0$ is the coordinate of the composite state.

The auxiliary monodromy matrix for this model can be parametrized as follows
\begin{equation}
T(u)=L_n(u)L_{n-1}(u)\ldots L_1(u)=\left(
\begin{array}{cc}
A(u) & B(u) \\
C(u) & D(u)
\end{array}
\right) \;,  \label{matmon}
\end{equation}
where $n$ is the number of the reggeized gluons. The transfer matrix is the
trace of the monodromy matrix

\[
t(u)=A(u)+D(u)=\sum _{j=0}^n Q_j \,u^{n-j}\,,
\]
where
\begin{equation}
Q_j=\sum _{i_1>i_2>...>i_j}p_{i_1}p_{i_2}...p_{i_j}\,\rho
_{i_1i_2}...\rho _{i_{j-1}i_j} \rho _{i_ji_1}\,. \label{intmov}
\end{equation}
The eigenvalues $ \Lambda (u) $ of $t(u)$ take the form
\[
\Lambda (u)=2\,u^n+q_2\,u^{n-2}+q_3\,u^{n-
3}+\ldots+q_n\,.
\]
where  $q_j,\;2\leq j\leq n$ are  eigenvalues of the integrals
of motion $Q_j$ \cite{lip2}. In particular, $q_2=-m(m-1)$ is the
eigenvalue of the
holomorphic Casimir operator (\ref{casi}) and $m$ is the conformal
weight.

\bigskip

The operator $C(u)$ annihilates the pseudovacuum state for the $s=-1$
model \cite{fk},
\begin{equation}
C(u)\,\Omega_ 0 =0 \quad ,\quad \Omega _0 =\prod _{r=1}^n \,\rho _{r0}^{-2}\,.
\end{equation}

The operators $ B(u) $ can be obtained directly from eq.(\ref{matmon}). We
find for $n=2$ and $n=3$,
\begin{eqnarray}
B^{(n=2)}(u) &=& -u \, (p_1+p_2) + p_1\, p_2 \; \rho_{12} \\ \cr
B^{(n=3)}(u) &=&-u^2 \, (p_1+p_2+p_3)+ u \, (p_1\, p_2 \; \rho_{12}+ p_1\,
p_3 \; \rho_{13} +p_2\, p_3 \; \rho_{23}) - p_1\, p_2\, p_3 \; \rho_{12}\;
\rho_{23} \nonumber
\end{eqnarray}
For arbitrary $n$ one obtains,
\begin{eqnarray}
B^{(n)}(u) &=&-\sum_{k=0}^{n-1} b_k \; u^{n-1-k} \quad {\mbox{where}}\quad
b_0 = P \equiv \sum_{i=1}^n p_i \quad , \quad b_1 = - \sum_{1\leq i < j \leq
n} p_i \; p_j \; \rho_{ij} \; , \cr \cr b_2 &=& \sum_{1\leq i_1 < i_2 < i_3
\leq n} p_{i_1} \; p_{i_2} \; p_{i_3} \; \rho_{i_1 i_2}\; \rho_{i_2
i_3}\quad , \cr \cr &&\ldots , \cr \cr b_l &=& (-1)^l \sum_{1\leq i_1 < i_2
< \ldots  <i_{l+1} \leq n} p_{i_1} \; p_{i_2} \; \ldots \; p_{i_l}\;
p_{i_{l+1}} \; \rho_{i_1 i_2}\; \rho_{i_2 i_3}\; \ldots \;\rho_{i_l i_{l+1}}
\; , \cr \cr &&\ldots , \cr \cr b_{n-1} &=& (-1)^{n-1} p_1\, p_2\, \ldots \,
p_n \; \rho_{12}\; \rho_{23} \, \ldots \, \rho_{n-1,n} \; . \label{bllio}
\end{eqnarray}

The operators $B$ with different spectral parameters commute

\[
\left[ B(u)\,,\,B(v)\right] =0\,
\]
and therefore, one can write them in factorized form as a product of
the operator zeros $\widehat{\lambda }_k $ of $B(u)$:
\[
B(u)=-P\,\prod_{k=1}^{n-1}\left( u-\widehat{\lambda }_k\right)
\quad , \quad \left[
\widehat{\lambda }_{k_1},\widehat{\lambda }_{k_2}\right] =\left[ \widehat{%
\lambda }_k,P\right] =0\,,
\]
following E. K. Sklyanin \cite{skl}.

\bigskip

The wave function describing composite states of reggeized gluons
in the holomorphic impact parameter space $\rho $ can be written as
follows \cite{skl} (see also \cite{fk})
\begin{equation}
\psi (\rho _1,\rho _2,\ldots,\rho _n;\rho _0)=Q\left( \widehat{\lambda }
_1 \right) Q\left( \widehat{\lambda }_2\right) \ldots \,Q\left(
\widehat{\lambda }_{n-1}\right) \prod_{r=1}^n\rho _{r0}^{-2} \quad ,\,
\quad \rho_{r0}=\rho _r-\rho _0 \; .
\end{equation}
The function $Q\left( \lambda \right) $ satisfies the Baxter equation
\cite{bax} 
\begin{equation}  \label{eqBa}
\Lambda (\lambda )\,Q\left( \lambda \right) =(\lambda +i)^n \; Q\left(
\lambda +i\right) +(\lambda -i)^n \; Q\left( \lambda -i\right) \; .
\end{equation}

For the Odderon case, the dependence of the energy from the
eigenvalues of the integrals of motion has been found with the
use of the Baxter equation \cite{YW} and of the duality symmetry
\cite{dual}.

In this paper we systematically develop the construction of composite
Reggeon states using the Baxter-Sklyanin (BS) representation,
in which the operator zeros $\widehat{\lambda }_k $ of
$B(u)$ are diagonal. The matrix elements relating the momentum and
BS representations obey solvable  ordinary differential
equations for $n=2,\, 3$. These matrix elements are elementary
functions for the pomeron case and  hypergeometric functions for the
odderon case. In the BS representation the wave function of the composite state
is written as a product of the Baxter functions and the
pseudo-vacuum state.

For the pomeron, we provide general formulas for the Baxter function
valid in the whole complex $ \lambda $ plane and study its analytic
properties. It turns out that the most efficient way to solve the
Baxter equation in the present context is to use the pole expansions
(Mittag-L\"offler).

We show that the Pomeron wave function has no
singularities on the real axis as a function of $ \sigma=Re
\,\lambda$ and hence it can be
normalized. This corresponds to the single-valuedness condition in
the coordinate representation.

We derive also the analytic Bethe Ansatz equations and construct the
Baxter function as an infinite product of Bethe Ansatz roots.

The solution $Q(u)$ of the Baxter equation for the general
$n$-reggeon case is constructed as an infinite sum over poles of the
orders
from $1$ up to $n-1$. Their residues satisfy simple recurence relations.
It is shown, that the quantization condition for the integrals of
motion follows from the condition of the cancellation of
the pole singularities in the two
dimensional $\overrightarrow{\lambda}$-plane for the bilinear
combination of holomorphic and anti-holomorphic Baxter functions
$Q(\overrightarrow{\lambda})$ and the physical requirement, that
all Baxter functions with the same integrals of motion
yield the same energy.

For the odderon, we explicitly construct the  BS
representation and investigate the properties of the odderon wave
functions in this representation. The completeness and
orthogonality relations for these functions are discussed.

We derive new formulas for the eigenvalues of the Reggeon hamiltonian
written through the Baxter function. These formulas generalize the
result for the  Pomeron to any number of Reggeons. The energy turns to be
expressed in terms of the behavior of the Baxter function near its
poles at $ \lambda = i $ which are present for arbitrary $n$.

The BS representation promises to
be an appropriate starting point to find new composite Reggeon
states for $n > 3$. In particular, it will be interesting to
generalize the Odderon solution constructed in ref.\cite{BLV} to the case
of many Reggeons.

\section{BS representation for the wave function}

In order to solve the Baxter equation, one should fix the class of
 functions in which the solution is searched. The case of integer
 conformal weight $ m $ has been considered in
 refs.\cite{fk,kor,maa}. It was assumed there that the
solutions were entire functions with the asymptotics
\[
Q(\lambda )\sim \lambda ^{m-n} \,,
\]
but such functions do not exist for physical values of the conformal weights
$m$.

We want to find the conditions which should be satisfied by the solutions
of the Baxter equation from the known information about the eigenfunctions
$\Phi$ of the Schr\"odinger equation in the two-dimensional impact parameter
space $\overrightarrow{\rho} $ \cite{norm,muchos}. For
this purpose we perform an unitary transformation of the wave function
$\Phi$  to the BS representation
in which the operator $B(u)$ is diagonal.

To begin with, let us go to the momentum representation (with removed gluon
propagators):

\[
\Psi _{m,\widetilde{m}}(\overrightarrow{p_1},\,\overrightarrow{p_2},\,\ldots,%
\overrightarrow{p_n})=
\]
\begin{equation}  \label{impuls}
\prod_{r=1}^n\left( \overrightarrow{p_r}\right) ^2\int
\prod_{k=1}^n\left[ \frac{d^2\rho _k}{2\pi }\,\exp (i\overrightarrow{p_k}%
\cdot \overrightarrow{\rho _{k0}})\right] \,\Phi _{m,\widetilde{m}}(%
\overrightarrow{\rho _1},\,\overrightarrow{\rho _2},\,\ldots,\overrightarrow{%
\rho _n};\overrightarrow{\rho _0})\,.
\end{equation}
Here $\Phi _{m,\widetilde{m}}(\overrightarrow{\rho _1},\,\overrightarrow{%
\rho _2},\,\ldots,\overrightarrow{\rho _n};\overrightarrow{\rho _0})$ is the
wave function of the composite state in the two-dimensional impact parameter
space $\overrightarrow{\rho }$. It belongs to the principal series of the
unitary representations of the M\"obius group and is an eigenfunction of
its
Casimir operators

\[
\vec M^2\Phi _{m,\widetilde{m}}=m(m-1)\Phi _{m,\widetilde{m}}\quad ,\quad
\left( \vec M^{*}\right) ^2\Phi
_{m,\widetilde{m}}=\widetilde{m}(\widetilde{m%
}-1)\Phi _{m,\widetilde{m}}\,.
\]
Here

\[
m=\frac 12+i\nu +\frac n2\quad ,\quad \widetilde{m}=\frac 12+i\nu -\frac n2
\]
are conformal weights (the quantities $\nu $ and $n$ are correspondingly
real and integer numbers for the principal series of the unitary
representations). The Casimir operators of the M\"obius group are given by
eq.(\ref{casi}).

For example, for the pomeron and odderon we have respectively
\cite{norm, lip1},

\[
\Phi _{m,\widetilde{m}}^{(2)}(\overrightarrow{\rho _1},\,\overrightarrow{
\rho _2}\,;\overrightarrow{\rho}_0)=\left( \frac{\rho _{12}}{\rho _{10}\rho
_{20}}\right) ^m\left( \frac{\rho _{12}^{*}}{\rho _{10}^{*}\rho _{20}^{*}}%
\right) ^{\widetilde{m}}\; ,
\]

\[
\Phi _{m,\widetilde{m}}^{(3)}(\overrightarrow{\rho _1},\,\overrightarrow{%
\rho _2},\,\overrightarrow{\rho
_3}\,;\overrightarrow{\rho}_0)=\left( \frac{%
\rho _{23}}{\rho _{20}\rho _{30}}\right) ^m\left( \frac{\rho _{23}^{*}}{\rho
_{20}^{*}\rho _{30}^{*}}\right) ^{\widetilde{m}}\phi _{m,\widetilde{m}%
}(x\,,x^{*})\; ,
\]
where $\phi $ is a function of the anharmonic ratio,
\[
x=\frac{\rho _{12}\rho _{30}}{\rho _{10}\rho _{32}}\; .
\]
Due to the identity
\[
\sum _{k=1}^n p_k \, \rho _k =\frac{P}{n}\, \sum _{k=1}^n \rho _k
\,+\,\sum _{k=1}^{n-1} \rho _{k,k+1}\,
\sum _{r=1}^k \left( p_r-\frac{P}{n} \right)\,,\,\,P=\sum _{k=1}^n p_k\,,
\]
we can express the quantity $B(u)$ in momentum representation in
terms of $P,\, \,p_k-\frac{P}{n}$ and the operators
\[
\rho _{k,k+1}=-i\left( \frac \partial {\partial p_k}-\frac \partial
{\partial p_{k+1}}\right)=
-i\frac{\partial}{ \partial \left(\sum _{r=1}^k p_r \right)} \,,\,\,\,
k=1,\, 2, \,\ldots,\,n-1 \,\,.
\]
It is convenient to introduce the new independent variables
\be \label{deftk}
P=\sum_{k=1}^np_k\,,\,\,t_1=\ln \frac{p_1}{P-p_1}\,,\,\,t_2=\ln
\frac{p_1+p_2
}{P-p_1-p_2}\, , \ldots ,\, t_{n-1}=\ln \frac{P-p_n}{p_n} \; .
\ee
The quantities $t_k$ take their values in a strip of the complex
plane
\[
-\infty <Re\,t_k<\infty \,,\,\,\,-\pi <Im\,t_k<\pi \; .
\]

There is a helpful representation for the operators $b_l$ besides
that
given by eq. (\ref{bllio}). Namely,
\begin{equation}
b_l=(-1)^l \sum _{1\leq i_1<i_2<\ldots<i_{l}\leq n}
\left(\sum _{r_1=1}^{i_1}p_{r_1} \right) \left( \sum
_{r_2=i_1+1}^{i_2}p_{r_2} \right)
\ldots \left( \sum _{r_{l+1}=i_l+1}^{n}p_{r_{l+1}}\right) \,\,
\prod _{s=1}^l \rho _{i_s,i_s+1}\,, \label{b1}
\end{equation}
which is related with  the duality transformation \cite{dual} consisting
of the cyclic permutation
\[
p_k \rightarrow \rho _{k,k+1} \rightarrow p_{k+1}
\]
and the transposition of the operator multiplication.
Note, that with the use of the duality symmetry the
wave function in the momentum space (\ref{impuls}) for $P=0$ is
proportional to the same function in the coordinate space
\cite{dual}. Furthermore, the function for arbitrary $P$ can be
obtained by an appropriate M\"obius transformation.

In the variables $t_1, \ldots , t_{n-1}$ the matrix element $B(u)$ takes
the following
form
\[
B(u)=-\sum_{k=0}^{n-1}b_k\,u^{n-1-k}\,,
\]
where the operators $b_k$ are given by
\[
b_k=P\,\sum_{1\leq l_1<l_2<\ldots<l_k\leq n-1}\,\prod_{r=1}^{k-1}\left(
1-e^{t_{l_r}-t_{l_{r+1}}}\right) \,\prod_{s=1}^ki\,\frac \partial {\partial
t_{l_s}}\,.
\]
This representation  can be obtained from  eq.(\ref{b1}) for $b_l$
taking into account the formulas:
\[
i\frac{\partial }{\partial t_k}=i
\frac{(\sum _{r=1}^k p_r)\,(P-\sum _{r=1}^k p_r)}{P}
\frac{\partial }{\partial (\sum _{r=1}^k p_r )}=
-\frac{(\sum _{r=1}^k p_r)\,(P-\sum _{r=1}^k p_r)}{P}
\,\rho_{k,k+1}\,,
\]
\[
1-e^{t_{l_r}-t_{l_{r+1}}}=\frac{P \,(\sum _{s=1}^{l_{r+1}}p_s)}{(\sum
_{s=1}^{l_{r+1}}p_s) (P-\sum _{s=1}^{l_r}p_s)}\,.
\]

In particular, for $n=2,\,3,\,4$ we obtain

\[
B^{(2)}(u)=-P\left( u+i\frac \partial {\partial t_1}\right) ,
\]
\[
B^{(3)}(u)=-P\left[ u^2+iu\left( \frac \partial {\partial t_1}+\frac
\partial {\partial t_2}\right) -\left( 1-e^{t_1-t_2}\right) \frac \partial
{\partial t_1}\frac \partial {\partial t_2}\right] ,
\]

\begin{eqnarray}
B^{(4)}(u)=-P\left[ u^3+i\,u^2\sum_{r=1}^3\partial _r-u\sum_{1\leq
l_1<l_2\leq 3}\left( 1-e^{t_{l_1l_2}}\right) \partial _{l_1}{\partial }%
_{l_2}-i\prod_{k=1}^2\left( 1-e^{t_{k,k+1}}\right) \partial _1\partial
_2\partial _3\right] ,  \label{bennuv}
\end{eqnarray}
where

\[
t_{l_1l_2}=t_{l_1}-t_{l_2}\,,\,\,\partial _r=\frac \partial {\partial
\,t_r}\,.
\]
Since in the momentum representation the norm of the wave function is given
by
\[
\left\| \Psi _{m,\widetilde{m}}\right\| ^2=\int \prod_{r=1}^n
\frac{d^2p_r}{%
\left| p_r\right| ^2} \; \left| \Psi
_{m,\widetilde{m}}(\overrightarrow{p_1}%
,\, \overrightarrow{p_2},\,\ldots,\overrightarrow{p_n})\right| ^2\,,
\]
we obtain after extracting the factor $\delta ^2(P-\sum _{k=1}^n
p_k)$ from
$\Psi _{m,\widetilde{m}}$
in the new variables $t_1, t_1^*;\ldots ; t_{n-1}, t_{n-1}^* $

\[
\left\| \Psi _{m,\widetilde{m}}\right\| ^2= \int
\prod_{r=1}^{n-1}d^2t_r\prod_{s=1}^{n-2}\left| 1-e^{t_s-t_{s+1}}\right|
^{-2}\left| \Psi _{m,\widetilde{m}}\right| ^2\,.
\]
The operators $B(u)$ are symmetric $B=B^t$ with respect to this norm
with the
weight

\[
\prod_{s=1}^{n-2}\left| 1-e^{t_s-t_{s+1}}\right| ^{-2}.
\]

The eigenvalues of the operator zeroes $\widehat{\lambda }_k$ and
$\widehat{\lambda^* }_k $ of $B(u)$ in the holomorphic and anti-holomorphic
space have the form,
\[
\lambda _k=\sigma _k+i\frac{N_k}2\quad , \quad \lambda_k^* =\sigma _k-i
\frac{N_k}2
\]
where $\sigma _k$ is real and $N_k$ is integer. The unitary transformation
between $t$- and $\lambda $- representations conserves the norm of the
wave function
\[
\left\| \Psi _{m,\widetilde{m}}\right\| ^2=
\prod_{r=1}^{n-1} \left( \int_{-\infty }^{+\infty} d\sigma _r\sum_{N_r=-\infty
}^{+\infty} \right) \,\left| \Psi _{m,\widetilde{m}}\right| ^2.
\]

Let us introduce the kernel
$U_{\overrightarrow{\lambda _1}, \ldots, \overrightarrow{\lambda _{n-1}}}
(\overrightarrow{t_1}, \overrightarrow{t_2}, \ldots ,
\overrightarrow{t_{n-1}}) $ for the
unitary transformation between the  $t$- and $\lambda$-
representations. It satisfies the eigenvalue equations
\begin{eqnarray}
B(u)\,U_{\overrightarrow{\lambda _1}, \ldots, \overrightarrow{\lambda _{n-1}}}
(\overrightarrow{t_1},  \ldots ,
\overrightarrow{t_{n-1}})&=& -P\,\prod
_{k=1}^{n-1}(u-\lambda_k^*)\,U_{\overrightarrow{\lambda _1},
\ldots, \overrightarrow{\lambda _{n-1}}}
(\overrightarrow{t_1},  \ldots ,
\overrightarrow{t_{n-1}}) \,,\cr \cr
B(u)^*\,U_{\overrightarrow{\lambda _1}, \ldots, \overrightarrow{\lambda
_{n-1}}}
(\overrightarrow{t_1}, \ldots ,
\overrightarrow{t_{n-1}})&=& -P\,\prod _{k=1}^{n-1}(u-\lambda
_k)\,U_{\overrightarrow{\lambda _1},
\ldots, \overrightarrow{\lambda _{n-1}}}
(\overrightarrow{t_1}, \ldots ,
\overrightarrow{t_{n-1}})
\end{eqnarray}
and the orthogonality relations
\bea
&&\int \prod _{k=1}^{n-1} \frac{d^2\,t_k}{(2\pi )^2}\,\,\prod
_{r=1}^{n-2}|1-e^{t_r-t_{r+1}}|^{-2} \; U_{\overrightarrow{\lambda
_1}, \ldots, \overrightarrow{\lambda _{n-1}}} \; U^*_{\overrightarrow{\lambda'
_1}, \ldots, \overrightarrow{\lambda' _{n-1}}}= \cr \cr
&& \Sigma_P \prod _{t=1}^{n-1} \left[ \delta
(\sigma
_k-\sigma' _{r_k}) \; \delta _{N_k,N'_{r_k}} \right] \; .
\eea
Here, $ \lambda_k^* , \; \lambda_k $ stands for the eigenvalues of the
operators $ \widehat{\lambda }_k $ and  $ \widehat{\lambda }_k^* $,
respectively and the symbol $\Sigma_P$ means the sum over all possible
permutations $r_1,...,r_{n-1}$ of the indices $1,2,...,n-1$.

To construct the kernel  with the correct normalization, let us take
into account, that the Kronecker and Dirac $\delta $-functions appears
in the right hand side of the above equation as
a
result of the integration over the region
\[
t_2-t_1>>1, \,t_3-t_2>>1, \,\ldots,\,t_{n-1}-t_{n-2}>>1\,,
\]
corresponding to one of the two possibilities
\[
p_1<<p_2<<\ldots<<p_{n}
\]
or
\[
p_1>>p_2>>\ldots>>p_{n} \,.
\]
In this region the operator $B(u)$ simplifies as
\begin{equation}
B(u) \rightarrow -P \, \prod _{k=1}^{n-1}(u+i\partial_k)
\end{equation}
and therefore the kernel for the unitary transformation corresponds to
the Fourier transformation
\[
U_{\overrightarrow{\lambda _1}, \ldots, \overrightarrow{\lambda _{n-1}}}
(\overrightarrow{t_1}, \overrightarrow{t_2}, \ldots , \overrightarrow{t_{n-1}})
\rightarrow 2^{\frac{n-1}{2}} \Sigma_P \;
e^{i\Phi(\overrightarrow{\lambda_{r_1}}, \ldots,
\overrightarrow{\lambda_{r_{n-1}}})} \;
\exp [i\sum_{k=1}^{n-1}(t_k\lambda_{r_k}^*+t_k^*\lambda _{r_k})]\,,
\]
where $\Phi$ are some phases.

\section{BS representation for the pseudo-vacuum state}

The wave function of the pseudo-vacuum state in the momentum representation
is
\begin{equation} \label{estref}
\Phi_{m=n,\widetilde{m}=n}^0(\overrightarrow{p_1},\,\overrightarrow{p_2},\,
\ldots, \overrightarrow{p_n})=\int \prod_{k=1}^n\left[ \frac{4\;d^2\rho _k}{\pi
\,\left| \rho _{k0}\right| ^4}\;\exp (i\overrightarrow{p_k}\cdot
\overrightarrow{\rho _{k0}})\right] =\prod_{k=1}^n\left| p_k\right| ^2\ln
\left| p_k\right| ^2\,.
\end{equation}
where we subtracted from the distribution $\left| \rho \right| ^{-4}$ its
singular part \cite{gel}. This function is an eigenfunction of the
transfer
matrix
with the eigenvalue
\[
\Lambda (u) = (u+i)^n+(u-i)^n.
\]
In particular, its M\"obius conformal weights are
\[
m=\widetilde{m}=n
\]
and $ q_k = 0 $ for odd values of $ k $. Notice that the Baxter
function for the pseudo-vacuum state is $u$-independent.

It is important to know the wave function of the pseudo-vacuum state in the
BS representation. We see from eq.(\ref{bennuv}) that
$\widehat{\lambda }_1$ in the new variables for the Pomeron state
takes the simple form
\[
\widehat{\lambda }_1=-i\frac \partial {\partial t_1}
\]
and the change of the basis results in
\begin{equation}\label{cambase}
<p_1\,p_2|P,\lambda _1, \lambda _1^{*} >= \left| \frac{P}{p_1\,p_2}
\right|^2
\left( \frac{p_1}{p_2} \right)^{i\lambda _1^{*} }\;\left(
\frac{p_1^{*}}{p_2^{*}} \right)^{i\lambda _1}\;\delta^{(2)}
\left( P-p_1-p_2 \right)   \; . \label{baseN}
\end{equation}

One can obtain from eqs. (\ref{estref}) and (\ref{cambase}) the Pomeron
pseudo-vacuum wave function in the new variables as

\[
\Phi^0(\overrightarrow{P},\overrightarrow{\lambda _1})=\left|
P\right| ^4\int
d^2p\,\left( \frac p{1-p}\right) ^{-i\lambda _1^*}\left(
\frac{p^{*}}{1-p^{*}} \right)^{-i\lambda _1}
\ln \left| p\right| ^2\,\ln \left| 1-p\right| ^2.
\]
It can be written as follows

\[
\Phi ^0(\overrightarrow{P},\overrightarrow{\lambda _1})=\left|
P\right|
^4\lim_{\mu  \rightarrow \sigma}\frac \partial {\partial \sigma }\frac
\partial {\partial \mu }\int d^2p\,\left| p\right| ^{-2i\sigma }\left|
1-p\right| ^{2i\mu }\left[ \frac{p^{*}(1-p)}{p(1-p^*)}\right]^{\frac N2},
\]
where  $ \lambda _1=  \sigma +i N/2 $.
The integral is calculated in a closed form with the use of the
anti-Wick rotation (see next section):
\begin{equation}
\Phi ^0(\overrightarrow{P},\overrightarrow{\lambda })= 4\pi  (-1)^N \;
\left| \lambda \right|^2 \; \lim_{\mu  \rightarrow \sigma} {1 \over
(\mu - \sigma)^3 }.
\end{equation}
The fact that the wave function for the pomeron pseudo-vacuum state
turns out
to be
divergent is connected with the fact that such state having the
weights $m=
\widetilde{m}=2$ is outside the space of physical states. However it
is
important to  normalize correctly the Baxter function. The natural
regularization of the Heisenberg model can be provided by changing
the
spin representation $s
\rightarrow
-1+\epsilon$ without losing its integrability. It would lead in
particular to the
modification of the pseudo-vacuum state (\ref{estref}) and to the
convergence of integrals after their analytic continuation.

The above result for $\Phi
^0(\overrightarrow{P},\overrightarrow{\lambda })$
can be obtained in a simpler way by taking into account,
that in the integral transformation to the $\lambda $-representation the
large momenta $p$ dominate
\[
\Phi ^0(\overrightarrow{P},\overrightarrow{\lambda })\simeq (-1)^N\left|
P\right| ^4\int d^2p\,e^{-i\frac{\lambda ^{*}}p}e^{-i\frac \lambda
{p^{*}}}\ln
^2\left| p\right| ^2=(-1)^N\left| P\right| ^4\left| \lambda \right| ^2c,
\]
where the leading divergent contribution to $c$ does not depend on $%
\lambda $. It corresponds to the following
simplification of the operator $B(u)$ for $ p \gg P $,

\[
B^{(2)}(u)=-P\left( u-{p^2 \over P} \; i \, \frac \partial {\partial
p}\right) \,.
\]

In the case of three particles we have for large $ p_1 , \; p_2, \;  p_3
$ with fixed $P$,
\bea
B^{(3)}(u)&=&-P \left[ u^2+2iu\frac \partial {\partial t}-y\left(
\frac{\partial^2}{\partial y^2}-\frac{\partial^2}{\partial
t^2}\right)\right] \\ \cr
t&=&t_1 + t_2 = P\left( p_1^{-1}-p_3^{-1}\right)\ll 1  \quad  , \quad
y=t_1 - t_2 = P\left(p_1^{-1}+p_3^{-1}\right) \ll 1 \; . \nonumber
\eea
Its eigenvalues and eigenfunctions for small $t$ and $y$ are
$$
B^{(3)}(u)=-P(u-\lambda _1^*)(u-\lambda _2^*) \,,
$$
\be \label{tychic}
\varphi _{\lambda _1^*\lambda _2^*}^{1}(t,y)=
e^{\frac{i}{2}(\lambda ^*
_1+\lambda _2 ^*)t}\left[ 1 - \lambda_1^* \; \lambda_2^* \; y \; \ln
 \,y\right] \,,\,\,\varphi _{\lambda _1^*\lambda _2^*}^{2}(t,y)=
e^{\frac{i}{2}(\lambda ^*
_1+\lambda _2 ^*)t}\,y\,.
\ee
Imposing the property of the single-valuedness we can write the
transition amplitude in the two-dimensional space:
$$
U_{\overrightarrow{\lambda _1},\,\overrightarrow{\lambda _2}}
(\overrightarrow{t_1},\,\overrightarrow{t_2})=
c_{\overrightarrow{\lambda _1},\,\overrightarrow{\lambda _2}}\;
 e^{\frac{i}{2}\left[ (\lambda ^*_1+\lambda _2 ^*)t + (\lambda
_1+\lambda _2)t^*\right] }\, \left( \frac{y}{\lambda _1
\lambda _2}+ \frac{y^*}{\lambda _1^*
\lambda _2^*}-\ln \,|y|^2\right)\,,
$$
where the constant $c_{\overrightarrow{\lambda
_1},\,\overrightarrow{\lambda _2}}$ for the normalized function
$U$ is calculated below (see (\ref{oddass})).

However, the pseudo-vacuum state and  other solutions
with integer conformal weights do not belong to the
space of the physical states. Therefore their unitary
transformation to the BS representation should be special
and the normalization of the unitary transformation could
include only the integration over the large momenta. In this case it
is more natural to write for the kernel of this transformation
the following expression
$$
U_{\overrightarrow{\lambda _1},\,\overrightarrow{\lambda _2}}
(\overrightarrow{t_1},\,\overrightarrow{t_2})=
c^{ps}_{\overrightarrow{\lambda _1},\,\overrightarrow{\lambda _2}}\;
 e^{\frac{i}{2}\left[ (\lambda ^*_1+\lambda _2 ^*)t + (\lambda
_1+\lambda _2)t^*\right] }\; ,
$$
where the constant $c^{ps}_{\overrightarrow{\lambda
_1},\,\overrightarrow{\lambda _2}}$  does not
depend on $\lambda _{1,2}$.

Let us now consider the $n$-Reggeon case. Again, the large momenta
$\overrightarrow{p_k}, \; 1 \leq k \leq n$ presumably dominate the
pseudo-vacuum wave function $
\Phi^0(\overrightarrow{P},\,\overrightarrow{\lambda})$ when expressed
as integral transform of eq.(\ref{estref}). This large momenta regime
corresponds to small $ t_k , \; 1 \leq k \leq n-1$ according to
eq.(\ref{deftk}),
$$
Re \,t_k =\frac{P}{\sum_{r=1}^k p_r} + {\cal O}\left(P \over \sum_{r=1}^k
p_r \right)^2  \quad , \quad Im  \,t_k = \pi \; .
$$
Since the operator $B^{(n)}(u)$ for small $ Re \,t_k $ and $Im  \,t_k = \pi$
contains more derivatives $\partial _k$ than factors $t_{r_1r_2}$
compensating them, we obtain in this regime for an arbitrary number
$n$ of reggeons
$$
B^{(n)}(u) \simeq -P \left[ u^{n-1} + u^{n-2} i (n-1) \partial_t +
\ldots \right] \simeq  -P \left[ u^{n-1} -  u^{n-2} \sum_{j=1}^{n-1}
\lambda^*_j + \ldots \right]
$$
where $ t = t_1 + \ldots + t_{n-1} $.

Therefore, the transformation kernels for $n$ reggeons have the
large-$p$ behavior similar to the case of $ n = 2 $ and $3$,
$$
U_{\overrightarrow{\lambda _1},\ldots,\overrightarrow{\lambda
_{n-1}}}(\overrightarrow{t_1},\ldots,\overrightarrow{t_{n-1}})=
c^{ps}_{\overrightarrow{\lambda _1},\ldots,\overrightarrow{\lambda_{n-1}}}\, \;
\prod _{k=1}^{n-1}\left(
e^{i\frac{\sum _{l=1}^{n-1}\lambda ^*_l}{n-1}
\,t_k}\,e^{i\frac{\sum_{s=1}^{n-1}\lambda_s}{n-1} \, t_k^*}\right)
$$
where the constant $ c_{\overrightarrow{\lambda
_1},\ldots,\overrightarrow{\lambda_{n-1}}^{ps}} $ is fixed by the
normalization condition with the integration over the region of
large momenta.

Thus, the pseudo-vacuum state in the BS representation can be written
as follows,
$$
\Phi ^0 (\overrightarrow{P},_,\overrightarrow{\lambda
_1},\ldots,\overrightarrow{\lambda_{n-1}}) \sim |P|^{2n}  \;
c^{ps}_{\overrightarrow{\lambda_1},\ldots,\overrightarrow{\lambda
_{n-1}}}\, \times
$$
$$
\int \prod _{k=1}^{n-1}\left( d^2t_k
\,e^{i\frac{\sum _{l=1}^{n-1}\lambda ^*_l}{n-1}
\,t_k}\,e^{i\frac{\sum_{s=1}^{n-1}\lambda_s}{n-1}\, t_k^*}\right)\,
\prod_{r=1}^{n-2}|t_r-t_{r+1}|^{-2}
\,\prod_{m=1}^{n}|p_m|^2\ln{|p_m|^2} \sim
$$
$$
|P|^{2n}  \; c_{\overrightarrow{\lambda_1}, \ldots ,\overrightarrow{\lambda
_{n-1}}} \,\int \prod _{k=1}^{n-1}\left[
d^2p_k \; e^{\frac{i}{n-1}\left(
\frac{\sum _{l=1}^{n-1}\lambda ^*_l}{\sum _{r=1}^k p_r}+\frac{\sum
_{s=1}^{n-1}\lambda_s}{\sum _{r=1}^k p_r^*}\right)}\right]\,\prod
_{m=1}^{n}\ln{|p_m|^2} \,.
$$

Using dimensional arguments we can write the result of the
integration for the pseudo-vacuum wave function as
\begin{equation}\label{pseudo}
\Phi^0 (\overrightarrow{P},_,\overrightarrow{\lambda_1}, \ldots
,\overrightarrow{\lambda_{n-1}}) \sim |P|^{2n} \;
c^{ps}_{\overrightarrow{\lambda
_1}, \ldots ,\overrightarrow{\lambda_{n-1}}} \; \left|
\sum_{s=1}^{n-1}\lambda_s \right|^{2(n-1)}
\end{equation}
up to a divergent $\lambda$-independent factor which can be
regularized by changing the
value of the Heisenberg spins $s \rightarrow -1 + \epsilon$.

Thus, providing that only large momenta are essential in the unitary
transformation,
the pseudovacuum state in the BS representation is expressed
in terms of the normalization constant $c^{ps}_{\overrightarrow{\lambda
_1}, \ldots ,\overrightarrow{\lambda
_{n-1}}}$ times $ \left| \sum_{s=1}^{n-1}\lambda_s \right|^{2(n-1)} $.
For the pomeron $c_{\overrightarrow{\lambda}}$ does not depend on
$\lambda$ [see eq.(\ref{cambase})].
As it was argued above, for the kernel $U_{\overrightarrow{\lambda
_1},\ldots,\overrightarrow{\lambda_{n-1}}}
(\overrightarrow{t_1},\ldots,\overrightarrow{t_{n-1}})$, describing
the transition between the momentum and BS representations for the
pseudo-vacuum wave function, it is
natural to take into account in the normalization condition only
the contribution from large momenta (presumably this is valid
also for all
states with integer conformal weights $m$ and $\widetilde{m}$). We
obtain in this way,
\begin{equation}
c ^{ps}_{\overrightarrow{\lambda
_1}, \ldots ,\overrightarrow{\lambda
_{n-1}}}=1
\end{equation}
up to some $\lambda _k$ independent factor.

The above arguments
are
in accordance with the Sklyanin theory in which the pseudo-vacuum
state in the wave function is considered as a multiplier allowing to
write the other multiplier as a  product of the Baxter functions.
Because, as it will be shown below, the Baxter function for the
$n$-reggeon composite state contains the poles $Q(\lambda ) \sim
(\lambda -i \,r)^{-(n-1)}$ for $r=0,1,...$, it is natural to expect
that the wave function of the pseudo-vacuum state cancels some of these
poles. Moreover, since from each solution  of the Baxter equation
we can obtain other solutions  multiplying it by factors
$\sinh ^k (2 \pi  \,\lambda )$, this symmetry should appear as a
possibility to multiply the pseudo-vacuum state
by such factors. Generally, the pseudo-vacuum state
is not symmetric under the permutation of the parameters $\lambda
_1,\,\lambda _2, \,...,\lambda _{n-1}$. In order to find the Hamiltonian
in the Baxter-Sklyanin representation we show below that in the
region
where $\lambda _{n-1},\,\lambda ^*_{n-1} \rightarrow i$ and $\lambda _s
\rightarrow 0$
($s=1,2,...,n-2$) the holomorphic wave function has only a
single pole at $\lambda _{n-1}=i$. Therefore, in order to agree
with the Baxter representation we should substitute
\begin{equation} \label{psnorm}
\lim _{\lambda _{n-1},\,\lambda _{n-1} \rightarrow i}\,\,c^{ps}_{0, \ldots
,\,0,\overrightarrow{\lambda _{n-1}}} \,\rightarrow \,\sinh
^{n-2}(2\pi \,\lambda _{n-1}) \,\sinh
^{n-2}(2\pi \,\lambda ^*_{n-1}) \,.
\end{equation}

\section{BS wave function for the Pomeron}

The wave function of the Pomeron in the momentum representation is given by,
\begin{equation}\label{fdopom}
\Psi_{m,\widetilde{m}}(\overrightarrow{p_1},\,\overrightarrow{p_2}
)=\prod_{r=1}^2\left( \overrightarrow{p_r}\right)^2 \int \prod_{k=1}^2
\left[ \frac{d^2\rho_k}{2\pi } \,\exp
(i\overrightarrow{p_k}\overrightarrow{%
\rho _{k0}}) \right] \,\,\left( \frac{\rho _{12}}{\rho _{10}\rho _{20}}%
\right) ^m\left( \frac{\rho _{12}^{*}}{\rho _{10}^{*}\rho _{20}^{*}}\right)
^{\widetilde{m}}
\end{equation}
and corresponds to the contribution of a triangle diagram. According
to Appendix A the matrix element for this triangle diagram in the
momentum space is

\begin{equation}  \label{psipom}
\Psi_{m,\widetilde{m}}(\overrightarrow{p_1},\,\overrightarrow{p_2} )= C_{m,
{ \widetilde m}} \;
\; \int d^2 k \; \left[ {\frac{k \, (P-k)}{p_1-k
}}\right]^{{\widetilde m}-1} \; \left[
{\frac{k^*(P^*-k^*)}{p_1^*-k^*
}}\right]^{m-1} \; ,
\end{equation}
where
\[
C_{m, {\widetilde m}} = - (-1)^n\frac{i^{{\widetilde m}-m} \; m
\;{\widetilde m} }{2^{m+{\widetilde m} + 4} \; \pi^3}   \;
\frac{\Gamma(1-m)}{\Gamma({\widetilde m})} \; .
\]

Note, that the above integral over $\overrightarrow{k}$ is convergent at the
singular points of the integrand and at the infinity providing that $m$ and
$ \widetilde{m}$ correspond to the principal series of the unitary
representations

\[
m=\frac 12+i\nu +\frac n2\,,\,\,\widetilde{m}=\frac 12+i\nu -\frac
n2\; .
\]

Terms proportional to $ \delta^{(2)}(\overrightarrow{p_1})$ and
$\delta^{(2)}(\overrightarrow{p_2}) $ \cite{kkk} can be neglected
here since they are multiplied by $ \overrightarrow{p_1}^2 \;
\overrightarrow{p_2}^2 $.

One can analytically continue it for other values of $m$ and
$\widetilde{%
m }$. In particular, for $m=\widetilde{m}\rightarrow 2$, corresponding to
the pseudo-vacuum state, we obtain that the  leading
contributions from two regions: small $\left| p_1-k\right| $
\[
\int d^2k\,\left( \frac{k^{*}(p_1^{*}+p_2^{*}-k^{*})}{p_1^{*}-k^{*}}\right)
^{m-1}\,\left( \frac{k(p_1+p_2-k)}{p_1-k}\right) ^{\widetilde{m}-1}\simeq -%
\frac{2\pi \left| p_1\right| ^2\left| p_2\right| ^2}{m+\widetilde{m}-4}
\]
and large $\left| k\right| $
\[
\left| p_1\right| ^2\left| p_2\right| ^2\int d^2k\,\frac{\left[
k^{*}(p_1^{*}+p_2^{*}-k^{*})\right]^{m-1}}{(p_1^{*}-k^{*})^{m+1}}\,\frac{%
\left( k(p_1+p_2-k)\right) ^{\widetilde{m}-1}}{(p_1-k)^{\widetilde{m}+1}}
\simeq \frac{2\pi \left| p_1\right| ^2\left| p_2\right| ^2}{m+\widetilde{m}%
-4 }\,
\]
cancel. The final result turns out to be proportional to $\left| p_1\right|
^2\left| p_2\right| ^2\ln \left| p_1\right| ^2\ln \left| p_2\right| ^2$.

\bigskip

Thus, the Pomeron wave function $\Phi _{m,\widetilde{m}}(\overrightarrow{P}
,\,\overrightarrow{\lambda })$ in the BS representation is (for
$\lambda =-\lambda _1$)
\begin{eqnarray}\label{pomdiag}
&&\frac{\Phi
_{m,\widetilde{m}}(\overrightarrow{P},\,\overrightarrow{\lambda }
) }{P^{\widetilde{m}} \; (P^*)^m}=\int \frac{d^2p}{\left| p(1-p)\right| ^2}
\,\left( \frac p{1-p}\right) ^{i\lambda ^{*}}\left( \frac{p^{*}}{1-p^{*}}
\right) ^{i\lambda }\,\Psi _{m,\widetilde{m}}(\overrightarrow{p},\,
\overrightarrow{1}-\overrightarrow{p}) \cr \cr &&= C_{m,\widetilde{m}}\,\int
\frac{d^2p}{\left| p(1-p)\right| ^2}\,\left( \frac p{1-p}\right) ^{i\lambda
^{*}}\left( \frac{p^{*}}{1-p^{*}}\right) ^{i\lambda }\int
d^2 k\,\left[
\frac{k^{*}(1-k^{*})}{p^{*}-k^{*}}\right] ^{m-1}\,\left[ \frac{k(1-k)}{p-k}%
\right]^{\widetilde{m}-1} \; ,  \label{intpom}
\end{eqnarray}
where $ p=p_1/P, p^*=p_1^*/P^*, $ and $ k $ and $ k^* $ were also rescaled by $
P $ and $ P^* $, respectively. The integrand is a single-valued
function only for the following values of the variable $\lambda $:
\[
\lambda _1=  \sigma +i \, \frac N2 \quad ,\quad \lambda^*_1
=\sigma -i \,\frac N2
\]
where $\sigma $ and $N$ are correspondingly real and integer numbers.

Eq.(\ref{pomdiag}) admits a natural interpretation as the Feynman diagram:

\vskip 3 cm

\vertice

\vskip 3 cm

The Feynman rule is as follows: a line where a momentum $ p $ flows
has associated the (conformal) `propagator' $ \; p^{i\lambda^{*}} \;
(p^{*})^{i\lambda} $ and the anomalous dimensions for other lines are
linear functions of $m$ and $\widetilde{m}$.

To calculate $\Psi _{m,\widetilde{m}}(\overrightarrow{P},\,\overrightarrow{%
\lambda })$ we use the anti-Wick rotation $p_2 \to -ip_0$, $k_2=-ik_0$ and
introduce $i \epsilon $ to keep the singularities off the integration paths:
\begin{equation}  \label{antiW}
|p|^2 = p \; p^* \to p_1^2 -
p_0^2 -i \epsilon \; .
\end{equation}
Let us concentrate our attention on the integrals over $k^* $ and $p^* $.
The position of the singularities in $k^* $ and $p^* $ in the
integrand of eq.(\ref{intpom}):
\[
(p\, p^* -i \epsilon)^{i\lambda^* -1}\; \left[(1-p)\, (1-p^*) -i
\epsilon\right]^{-i\lambda -1} \, \left[(p-k)(p^*-k^*) -i
\epsilon\right]^{1-m}
\]
depend on the values of $k $ and $p $. Therefore, the three singularities in
$p^* $ (or $k^* $) may be on one side of the real axis or one of them on
one side
and two others in the other side. In the first case we can deform the
contour on $%
p^* $ (or $k^* $) and the integral vanishes. We obtain for the non-zero
contributions after enclosing contours of integration over $k^{*}$ and
$p^{*}$ around the singularities of the integrand
\[
\Phi_{m,\widetilde{m}}(\overrightarrow{P},\,\overrightarrow{\lambda })=P^{
\widetilde{m}}\,  (P^*)^m \,C_{m,\widetilde{m}} \; i \,\sinh (\pi
\lambda )\,\sin(\pi \widetilde{m})\,\Psi
_{m,\widetilde{m}}(\overrightarrow{\lambda })\,,
\]
where
\begin{eqnarray}  \label{intIyII}
&&\Psi _{m,\widetilde{m}}(\overrightarrow{\lambda }) = \int_0^1 dp \;
\frac{
p^{i\lambda^{*}-1}}{(1-p)^{1+i\lambda ^{*}}}\int_0^p dk \; \frac{(p-k)^{1-
\widetilde{m}}}{[k(1-k)]^{1-\widetilde{m}}} \int_1^\infty dp^{*} \; \frac{
(p^*)^{i\lambda -1}}{(p^{*}-1)^{1+i\lambda }} \int_{-\infty }^0 dk^{*}\;
\frac{
(p^{*}-k^{*})^{1-m}}{[k^{*}(k^{*}-1)]^{1-m}} \cr \cr &&- (-1)^n\int_0^1 dp
\; \frac{p^{i\lambda ^{*}-1}}{(1-p)^{1+i\lambda ^{*}}} \int_p^1 dk \;
\frac{
(k-p)^{1-\widetilde{m}}}{[k(1-k)]^{1-\widetilde{m}}} \int_{-\infty }^0
dp^{*} \; \frac{(-p^{*})^{i\lambda -1}}{(1-p^{*})^{1+i \lambda }}
\int_1^\infty dk^{*} \; \frac{(k^{*}-p^{*})^{1-m}}{[k^{*}(k^{*}-1)]^{1-m}}
\; .
\end{eqnarray}

The integrals over $p^{*}$ and $k^{*}$ as well as those over $p$ and $k$ can
be transformed using relations of the type
\begin{eqnarray}  \label{fi12}
\Phi_{\widetilde{m}}^{(1)}(p)&=&\int_p^1 dk \;
\frac{(k-p)^{1-\widetilde{m}}}{
[k(1-k)]^{1-\widetilde{m}}}=-\int_{-\infty }^0 dk \; \frac{(p-k)^{1-%
\widetilde{m}}}{[-k\,(1-k)]^{1-\widetilde{m}}}\,,
\cr \cr
\Phi_{\widetilde{m}}^{(2)}(p)&=&\int_0^p dk \;
\frac{(p-k)^{1-\widetilde{m}}}{%
[k(1-k)]^{1-\widetilde{m}}}=-\int_1^\infty dk \;
\frac{(k-p)^{1-\widetilde{m}%
}}{[k\,(k-1)]^{1-\widetilde{m}}} \; .
\end{eqnarray}
Each of the two terms in eq.(\ref{intIyII}) factorizes into holomorphic and
anti-holomorphic functions. We can thus write $\Psi _{m,\widetilde{m}}(%
\overrightarrow{\lambda }) $ as

\[
\Psi _{m,\widetilde{m}}(\overrightarrow{\lambda })=i\frac{\sinh (\pi \lambda
)}{\sin (\pi m)}\left[ C_{\widetilde{m}}^{(2)}(\lambda
^{*})\,C_m^{(2)}(\lambda )+(-1)^n\,C_{\widetilde{m}}^{(1)}(\lambda
^{*})\,C_m^{(1)}(\lambda )\right] \,.
\]
where,
\begin{eqnarray}  \label{funC}
C_{\widetilde{m}}^{(1)}(\lambda ^{*})&=&\int_0^1\frac{p^{i\lambda
^{*}-1}\,d\,p}{(1-p)^{1+i\lambda ^{*}}}\,\Phi _{\widetilde{m}}^{(1)}(p)=i
\frac{\sin (\pi \widetilde{m})}{\sinh (\pi \lambda ^{*})}\,\int_{-\infty }^0
\frac{d\,p\,(-p)^{i\lambda ^{*}-1}}{(1-p)^{1+i\lambda ^{*}}}\Phi _{%
\widetilde{m}}^{(2)}(p)\, , \cr \cr
C_{\widetilde{m}}^{(2)}(\lambda
^{*})&=&\int_0^1\frac{p^{i\lambda ^{*}-1}\,d\,p}{(1-p)^{1+i\lambda ^{*}}}%
\,\Phi _{\widetilde{m}}^{(2)}(p)=-i\frac{\sin (\pi \widetilde{m})}{\sinh (\pi
\lambda ^{*})}\,\int_1^\infty \frac{d\,p\,p^{i\lambda ^{*}-1}}{%
(p-1)^{1+i\lambda ^{*}}}\Phi _{\widetilde{m}}^{(1)}(p)\,. \nonumber \\
\end{eqnarray}
The functions $\,\Phi _{\widetilde{m}}^{(1)}(p)$ and $\,\Phi _{\widetilde{m}
}^{(2)}(p)$ are related with each other as follows
\[
\lim_{\varepsilon \rightarrow 0,\,p<0}\left[ \Phi _{\widetilde{m}%
}^{(1)}(p-i\varepsilon )-\Phi _{\widetilde{m}}^{(1)}(p+i\varepsilon )\right]
=-i\,\sin (\pi \widetilde{m})\,\Phi _{\widetilde{m}}^{(2)}(p)\,,
\]
\[
\lim_{\varepsilon \rightarrow 0,\,p>1}\left[ \Phi _{\widetilde{m}%
}^{(2)}(p-i\varepsilon )-\Phi _{\widetilde{m}}^{(2)}(p+i\varepsilon )\right]
=i\,\sin (\pi \widetilde{m})\,\Phi _{\widetilde{m}}^{(1)}(p)\,\, .
\]
Upon changing the integration variables
\[
p\rightarrow 1-p,\,k\rightarrow 1-k
\]
one verifies from eq.(\ref{funC}) that

\[
C_m^{(2)}(\lambda )=C_m^{(1)}(-\lambda )
\]
and therefore
\[
\frac{\Phi
_{m,\widetilde{m}}(\overrightarrow{P},\,\overrightarrow{\lambda }%
) }{P^{\widetilde{m}}P^{*m}\,C_{m,\widetilde{m}}}=-\sinh ^2(\pi \lambda
)\,\,\left[ C_{\widetilde{m}}^{(1)}(\lambda ^{*})\,C_m^{(1)}(\lambda
)+(-1)^n\,C_{\widetilde{m}}^{(1)}(-\lambda ^{*})\,C_m^{(1)}(-\lambda
)\right] \,.
\]

We use the limiting value for the pseudo-vacuum state $m=\widetilde{m}%
\rightarrow 2$,

\[
\Phi _{\widetilde{m}}^{(1)}(p)\rightarrow \frac{p(1-p)}{2-\widetilde{m}}%
\,,C_{\widetilde{m}}^{(1)}(\lambda ^{*})\rightarrow \frac{\pi \,\lambda
^{*}%
}{\sinh (\pi \lambda ^{*})(2-\widetilde{m})},\,C_{m,\widetilde{m}%
}\rightarrow \frac{1}{2\pi [2-\widetilde{m}]}
\]
to normalize our wave function.

We can write the result in terms of the product of the Baxter functions
$Q(\lambda ,\,m)$ and the pseudovacuum state:
\begin{equation}  \label{fiqq}
\frac{\Phi
_{m,\widetilde{m}}(\overrightarrow{P},\,\overrightarrow{\lambda }%
) }{P^{\widetilde{m}}P^{*m}\,C_{m,\widetilde{m}}}=-{\frac{ (-1)^N\,\left|
\lambda \right|^2 }{m \; \widetilde{m} }} \left[ Q(\lambda
^{*},\widetilde{m}%
)\,Q(\lambda ,m)+(-1)^n\,Q(-\lambda ^{*},\widetilde{m})\,Q(-\lambda
,m)\right] \, ,
\end{equation}
where  $Q(\lambda ,m)$ is defined as
\begin{eqnarray}  \label{intQ}
Q(\lambda ,m)&=&-m \frac{\sinh (\pi \lambda )}\lambda \int_0^1\frac{%
p^{i\lambda -1}\,d\,p}{(1-p)^{1+i\lambda }}\int_p^1\frac{(k-p)^{1-m}\,d\,k}
{
[k(1-k)]^{1-m}} \cr \cr &=&-i\sinh (\pi \lambda )\int_0^1\frac{p^{i\lambda
-1}\,d\,p}{ (1-p)^{1+i\lambda }}\int_p^1\frac{(k-p)^{-m}\,d\,k}{%
[k(1-k)]^{1-m}}\, .
\end{eqnarray}
These two equivalent integral forms of the Baxter function are related with
integrating by parts in $p$ and using the identity

\[
\frac d{dk}\left[ \frac{k(1-k)}{k-p}\right]^m=-m\,\frac{[k(1-k)]^{m-1}}{
(k-p)^{1+m}}\left[ p(1-p)+(k-p)^2\right] \; .
\]
This corresponds to the fact that the pomeron wave function is an
eigenfunction of the Casimir operator of the M\"obius group

\[
p(1-p)\frac{d^2}{dp^2}\,\Phi _{\widetilde{m}}^{(1,2)}(p)=m(1-m)\,\Phi _{%
\widetilde{m}}^{(1,2)}(p)\,.
\]

Using eq.(\ref{intQ}) and the identity
\[
\lambda \; p^{i\lambda-1 }\;(1-p)^{-i\lambda-1 } =- i \; {\frac{d }{dp}}
\left[ \; p^{i\lambda }\;(1-p)^{-i\lambda} \right]
\]
we obtain that
\[
\lambda ^2 \, Q(\lambda ,m)=-i \; m \; \sinh (\pi \lambda ) \int_0^1\frac{
p^{i\lambda }\,d\,p}{(1-p)^{i\lambda }}\,\frac d{dp}\int_p^1\frac{
(k-p)^{1-m}\,d\,k}{[k(1-k)]^{1-m}}
\]
and
\[
\left( \frac p{1-p}\right) ^{i(\lambda +i)}+2\left( \frac p{1-p}\right)
^{i\lambda }+\left( \frac p{1-p}\right) ^{i(\lambda -i)}=\frac
1{p(1-p)}\left( \frac p{1-p}\right) ^{i\lambda },
\]
The Baxter equation for the pomeron (\ref{eqBa}) follows using both
integral representations (\ref{intQ}) for $Q(\lambda ,m)$
\begin{equation}  \label{ecbpom}
(\lambda +i)^2Q(\lambda +i,m) - 2\lambda ^2 \, Q(\lambda ,m) +(\lambda
-i)^2Q(\lambda -i,m)= m(1-m) \,Q(\lambda ,m)\,.
\end{equation}

It should be noticed that if $Q(\lambda ,m) $ is a solution of the Baxter
equation for the Pomeron, then $Q(-\lambda ,m) $ is also a solution.

\section{Analytic properties of the Baxter function for the Pomeron}

The functions $\Phi _m^{(1)}(p)$ and $\Phi _m^{(2)}(p)$ defined by
eq.(\ref{fi12}) can be written in
terms of the hypergeometric function $_2F_1(\alpha, \beta ; \gamma ; z) $
and the Legendre function $P_{m-1} (z) $ \cite{gr}
\begin{eqnarray}
\Phi _m^{(2)}(p)&=&\Phi _m^{(1)}(1-p) \cr \cr &=& \frac{\pi \,(1-m)}{\sin
(\pi \,m)}\,p\,F(1-m,\,m;\,2\,;\,p)\cr \cr &=& {\frac{\pi }{m\,\sin (\pi
\,m)%
}}\,p(1-p)\frac{d}{d\,p} P_{m-1}(1-2p) \; .
\end{eqnarray}
Therefore, we have for the Baxter function
\begin{eqnarray}  \label{funba}
Q(\lambda ,m) &=& -i \; \frac{\pi \,\sinh (\pi \lambda )}{\sin (\pi \,m)}
\int_0^1 dp \; (1-p)^{i\lambda -1} \; p^{-i\lambda-1} \; P_{m-1}(1-2p) \\
\cr &=&-\frac{\pi^2 \, m \, (1-m)}{\sin \pi \,m }\,\,_3F_2(-i\lambda
+1,2-m,\,1+m;\,2,\,2;\,1)\; ,  \label{funba2}
\end{eqnarray}
where the generalized hypergeometric function $_3F_2$ is defined
as follows \cite{gr}
\[
_3F_2(\alpha _1,\,\alpha _2,\,\alpha _3\,;\,\beta _1,\,\beta
_2;\,z)=\sum_{k=0}^\infty \frac{\left( \alpha _1\right) _k\,\left( \alpha
_2\right) _k\,\left( \alpha _3\right) _k}{\left( \beta _1\right) _k\left(
\beta _2\right) _k}\,\frac{z^k}{k!}\,,\,\,\left( \alpha \right) _k=\frac{
\Gamma (\alpha +k)}{\Gamma (\alpha )}\,.
\]
One can find for the pseudo-vacuum state at $m \to 2 $ ,
\be \label{m2}
Q(\lambda ,m)  = \frac{2\pi }{m-2} + {\cal O}(1)
\; .
\ee
We thus have the following series representation for the Baxter function
\begin{equation}  \label{f32}
Q(\lambda ,m) = -\pi \sum_{k=1}^{\infty} {\frac{k \, \Gamma(m+k) \;
\Gamma(1-m+k) }{(k !)^3 }} \;
{\frac{\Gamma(k-i\lambda)}{\Gamma(1 -i\lambda)}} \; .
\end{equation}
The late terms in this series behave as $k^{-1 -i\lambda} $. Hence, this is
a convergent series for  $ Im \,\lambda < 0$.

In order to analytically continue the Baxter function to the upper $\lambda$
plane we insert in eq.(\ref{funba}) the series representation of the
Legendre function \cite{leb}
\begin{eqnarray}  \label{legen}
P_{m-1}(1-2p) &=& {\frac{\sin^2 \pi m }{\pi^2}} \sum_{k=0}^{\infty} {\frac{
\Gamma(m+k) \; \Gamma(1-m+k) }{(k !)^2}} \left[ 2 \, \psi(k+1)
\right. \cr \cr
&-& \left.\psi(k+1-m) -\psi(k+m)- \ln(1-p) \right](1-p)^k
\end{eqnarray}
Integrating term by term in eq.(\ref{funba}) yields the series,
\begin{eqnarray}  \label{qserie}
&&Q(\lambda ,m) =  {\frac{i \, \pi }{\lambda}} +
{\frac{
\sin \pi m }{\Gamma (1+i \lambda )}} \;
\sum_{k=1}^{\infty}
{\frac{ k \; \Gamma(m+k) \; \Gamma(1-m+k) }{(k !)^3}} \;
\Gamma(k+i\lambda)\times \cr \cr
&&\left[ 2 \, \psi(k+1) + \psi(k) -\psi(k+1-m)
-\psi(k+m)- \psi(i\lambda + k) \right]
\end{eqnarray}
The late terms in this series behave as $k^{-2 +i\lambda} $. Hence, this is
a convergent series for $ Im \,\lambda >-1 $.
Eq.(\ref
{qserie}) explicitly display simple poles at
\[
\lambda = 0, i, 2i, 3i, \ldots, l \, i, \ldots\; .
\]
Actually, the poles at $\lambda = +il$ ($l=1,2, \ldots$) arise from the
logarithmic singularity of $P_{m-1}(1-2p) $ near $p = 1 $ in the integral
(\ref
{funba}) [see eq.(\ref{legen})].

Direct calculation yields from eq.(\ref{qserie}) for the residue at these
points
\begin{eqnarray}  \label{resi}
r_l(m) &\equiv& \lim_{\lambda \rightarrow il} \; [\lambda - i l ]\;
Q(\lambda ,m) \cr \cr r_0(m) &=& i \, \pi \\ \cr r_l(m)&=&-i \, \pi\, m
(1-m) \, _3F_2(-l +1,2-m,\,1+m;\,2,\,2;\,1) = -{\frac{ \sin \pi m
}{i \,
\pi} } \; Q(-il ,m) \nonumber
\end{eqnarray}
for $l=1,\,2, \ldots $. It is interesting to notice that the residues of
$Q(\lambda ,m)$ at $\lambda
= +il$ ($l=1,2, \ldots$) are expressed in terms of $Q(\lambda ,m)$ at
$\lambda = -il$ ($l=1,2,\ldots$).

In summary, eqs.(\ref{f32}) and (\ref{qserie}) explicitly show that $%
Q(\lambda ,m)$ is a meromorphic function of $\lambda $ with simple poles at
$%
\lambda = +il $ ($l=0,1,2,\ldots$). The appearance of these poles is
related with the logarithmic singularities of the wave function at
$p_1=0$ and $p_2=0$.

The pomeron wave function considered as a function of
real $\sigma$ for even $N$ may have in principle singularities
[Recall that $ \lambda = \sigma + i \frac{N}2 $]. However, we find from
eq.(\ref{fiqq}) for non-vanishing $N$,
\bea
&&\mbox{Pole~~at}\; \sigma=0 \; \mbox{of} \; \left[
Q(\lambda^{*},\widetilde{m})\,Q(\lambda,m)+(-1)^n\,Q(-\lambda
^{*},\widetilde{m})\,Q(-\lambda ,m)\right] =\cr \cr
&&={1 \over \sigma}\left[ r_l(m) \;  Q(-il ,\widetilde{m}) - (-1)^n
Q(-il ,m) r_l(\widetilde{m}) \right] = 0
\eea
where $ l = N/2 $. Here we used eq.(\ref{resi}) and the relation
$$
\sin \pi m =  (-1)^n \; \sin \pi\widetilde{m} \; .
$$
It is important to notice that in the wave function the pole at
$\sigma=0$ and $N=0$ is also cancelled by the factor
corresponding to the pseudo-vacuum state, thus
allowing to normalize the pomeron eigenfunctions.
Analogous cancellations  of poles of the Baxter
function at real $\sigma $ for a higher number of Reggeons  lead to
the quantization of the integrals of motion $ q_r , \; r > 2 $, as it
will shown below.
In coordinate space this quantization
appears as a consequence of the single-valuedness condition (see
\cite{YW}).

\bigskip

Notice that both eqs.(\ref{f32}) and (\ref{qserie}) exhibit the $m
\Longleftrightarrow 1-m $ symmetry:
\[
Q(\lambda ,m)=Q(\lambda ,1-m)\; .
\]

Therefore, $Q(\lambda ,m)$ depends on $m $ through the invariant combination
$m(1-m) $ as we check explicitly below [see
eqs.(\ref{relrec})-(\ref{relrec2}%
)]. Note, that this combination for the principal series takes the form
\be \label{munom}
m(1-m) = \frac14 -  \left(i\nu  + {n \over 2} \right)^2 .
\ee

We see from eq.(\ref{funba}) and (\ref{qserie}) that the function $Q(\lambda
,m)$ obeys the relation
\[
{\bar Q}(\lambda ,\bar m) = Q(-{\bar \lambda},m)\; .
\]
That is, $Q(\lambda ,m)$ is real for purely imaginary $\lambda $ and
real $ m $. We have for real $\lambda $ and  $ m $,
\[
\mbox{Re}\, Q(-\lambda ,m) = +\mbox{Re} \, Q(\lambda ,m) \quad , \quad%
\mbox{Im} \, Q(-\lambda ,m) = -\mbox{Im} \, Q(\lambda ,m)\; .
\]

\bigskip

The asymptotic behavior of  $ Q(\lambda ,m) $ for large $ \lambda $ is
derived in appendix B starting from the integral representation
(\ref{funba}),
\begin{eqnarray}  \label{asintQ1}
&&Q(\lambda ,m)  = 4 \, \sqrt{\pi} \; \left[
\left(4 \, i \; \lambda \right)^{m-2} \; \frac{ \Gamma\left(m -
\frac12 \right) \; \Gamma\left(2-m \right) }{\Gamma\left(m \right)
} + \right.  \\ \cr
&&+  \left. \left( 4 \, i \; \lambda\right)^{-1-m}
\; \tan \pi m \; \frac{ \Gamma\left(m \right) \; \Gamma\left(m+1
\right) }{\Gamma\left(m + \frac12 \right)} +
{\cal O}\left(\lambda^{m-4} , \lambda^{-m-3}\right) \right]
\; . \nonumber
\end{eqnarray}

\bigskip

The Baxter equation for the pomeron (\ref{ecbpom}) written in the form
$$
Q(\lambda ,m) = { (\lambda +i)^2 Q(\lambda +i,m) + (\lambda
-i)^2Q(\lambda -i,m) \over 2\lambda ^2 +  m(1-m) }
$$
would seem to suggest that $ Q(\lambda ,m) $ has singularities at the
zeros $ \pm i \; \eta_m $ of the denominator where
$$
\eta_m \equiv  \sqrt{\frac12 \; m(1-m)} = \sqrt{\frac18 -\frac12
\left(i \nu + {n \over 2} \right)^2 } \; .
$$
However, we know that $ Q(\lambda ,m) $ is analytic there. Therefore,
the following relation holds,
$$
{  Q(i \, \eta_m + i ,m) \over  Q(i \, \eta_m - i ,m)} = -
\left({\eta_m - 1 \over \eta_m + 1}\right)^2
$$

\subsection{Dispersion (Mittag-L\"offler) representation of the Baxter
function}

We obtain from the Baxter equation (\ref{ecbpom}) a
recurrence relation for the residues $r_l(m) $
\begin{eqnarray}  \label{relrec}
&&(l+1)^2 \; r_{l+1}(m) + (l-1)^2 \; r_{l-1}(m) = \left[ 2 \, l^2 + m(m-1)
\right] r_l(m) \quad \mbox{for} \; l \geq 1 \; , \cr \cr &&r_1(m) = m(m-1)
\; r_0(m)\,.
\end{eqnarray}
All residues are thus determined in terms of the residue at the origin $%
r_0(m) $. That is,
\begin{eqnarray}  \label{relrec2}
r_1(m) = m(m-1) \; r_0(m) \; , \; r_2(m) = \frac14 \; m(m-1) \left[2+
m(m-1)\right] \; r_0(m) \; , \cr \cr r_3(m) = \frac19 \; m(m-1) \; \left[3+
\frac52 \; m(m-1)+ \frac14 \; m^2(m-1)^2 \right] \; r_0(m) \; .
\end{eqnarray}
It is easy to check that eqs.(\ref{f32}) and (\ref{resi}) agree with these
results.

The asymptotic behavior of the residues $r_l(m) $ follows from the
recursion relation (\ref{relrec}). We find, for large $l$
\begin{equation}  \label{asires}
r_l(m) = c(m) \; l^{m-2} + c(1-m) \; l^{-m-1}\,.
\end{equation}

Therefore, we can write the following Mittag-L\"offler (dispersion)
representation for the Baxter function
\begin{eqnarray}  \label{reprML1}
Q(\lambda ,m) &=& \sum_{l=0}^\infty \frac{r_l(m)}{\lambda - il} \cr \cr
&=&{\frac{i \, \pi }{\lambda }} - {\frac{ \sin \pi m }{i \, \pi}}
\sum_{l=1}^\infty \frac{Q(-il ,m)}{\lambda - il}\,.
\end{eqnarray}
The asymptotic behavior (\ref{asires}) guarantees the convergence of this
sum for $-1< \mbox{Re} \, m<2$.

\bigskip

In general, we have for $m < p + 2 $ where $p $ is a positive integer or
zero%
\cite{lav}
\[
Q(\lambda ,m) = F_p(\lambda,m) + \sum_{l=0}^\infty r_l(m)
\left[ {\frac{1 }{%
\lambda - i \, l}} - h_{l,p}(\lambda) \right] \; ,
\]
where,
\begin{eqnarray}
F_p(\lambda,m) &=& \sum_{k=0}^p {\frac{ (\lambda+i)^k }{k!}} \; Q^{(k)}(-i
,m) \; , \cr \cr h_{l,p}(\lambda)&=& \sum_{k=0}^p {\frac{ i^{1-k} }{%
(l+1)^{k+1}}}\; (\lambda+i)^k \; .
\end{eqnarray}

For example,
\[
F_0(\lambda,m) = Q(-i ,m) = -\pi^2 \, {\frac{m(1-m) }{\sin \pi m }} \quad ,
\quad
\]
\[
F_1(\lambda,m) =Q(-i ,m) + (\lambda+i) Q^{\prime}(-i ,m) = -\pi^2 \, {\frac{
m(1-m) }{\sin \pi m }} + i \pi (\lambda+i) \sum_{k=2}^{\infty} {\frac{
\Gamma(m+k) \; \Gamma(1-m+k) }{(k-1) \; (k !)^2}}
\]

We get from eq.(\ref{qserie}) for $Q(\lambda ,m)$ in the limit $\lambda
\rightarrow i$,
\begin{equation}  \label{qcercai}
\lim_{\lambda \rightarrow i}Q(\lambda ,m)=\pi \; m\, (1-m)\left[ -\frac{i}{
\lambda -i}+2-\psi (m)-\psi (1-m)+2\psi (1)\right]
\end{equation}
and therefore

\begin{equation}  \label{derli}
-i\lim_{\lambda \rightarrow i}\frac d{d\,\lambda }\ln Q(\lambda ,m)=\frac
i{\lambda -i}+2-\psi (m)-\psi (1-m)+2\psi (1) \; .
\end{equation}
We obtain for the other (independent) solution $Q(-\lambda , m)$ of the Baxter
equation,
\begin{equation}
-i\lim_{\lambda \rightarrow -i}\frac d{d\,\lambda }\ln Q(-\lambda
,m)=\frac
i{\lambda +i}-2+\psi (m)+\psi (1-m)-2\psi (1) \; .
\end{equation}

The behaviour of the Baxter function $ Q(\lambda ,m) $ near its
nearest pole $ \lambda = i $ can be also computed from the
Mittag-L\"offler expansion (\ref{reprML1}) with the result
\be
Q(\lambda ,m)\buildrel{ \lambda \to i}\over= {i \, \pi \, m\, (1-m)
\over \lambda -i}-ir_0(m)+i \sum_{l=2}^\infty \frac{r_l(m)}{l - 1}
\ee
Equating this result with eq.(\ref{qcercai}) yields the sum rule,
$$
-ir_0(m)+\sum_{l=2}^\infty \frac{r_l(m)}{l - 1} = i \, \pi
\, m\,
(1-m)\left[ \psi (m)+\psi (1-m)-2\psi (1)-2\right] \; .
$$

\bigskip

For large $ \lambda $ the Mittag-L\"offler series (\ref{reprML1}) is
dominated by its late terms. Notice that the sum of residues
\be\label{sum0}
\sum_{l=0}^\infty r_l(m) = 0
\ee
vanishes. The sum of late terms can be approximated by an
integral. Using eq.(\ref{asires}) we find
$$
Q(\lambda ,m)\buildrel{ \lambda \gg 1}\over= c(m) \; {\pi \over \sin
\pi m } \; (i \; \lambda)^{m-2} + (m
\Longleftrightarrow 1-m )
$$
in perfect agreement with eq.(\ref{asintQ1}).

\subsection{Infinite Product Representation of the Baxter Function}

As we have seen, the Baxter function $Q(\lambda ,m) $ is a meromorphic
function of $\lambda$ with simple poles at $\lambda = +il $ ($l=1,2,
\ldots$%
). Therefore, the function
\[
{\frac{ Q(\lambda ,m) }{\Gamma(i\lambda )}}
\]
is an entire function. Entire functions can be represented as infinite
products over their zeroes \cite{lav}.

Numerical study of the Baxter function $Q(\lambda ,m) $ in its
Mittag-L\"offler representation showed that all zeroes of $Q(\lambda ,m) $
are in the positive imaginary $\lambda$ axis for Re $m < 2$. We have in
addition $p$ real zeroes for $p+4 > $ Re $m > p + 2$ where $p $ is a
positive integer or zero.

For large $\lambda$ the imaginary zeroes $\lambda_k, \; k=1,2,\ldots $ are
equally spaced and follows for large $k$ the law:
\begin{equation}  \label{asiceros}
\lambda_k =i \left[ k +1 - m + {\cal O}\left(
\frac{1 }{k^a } \right) \right]
\end{equation}
for Re $m < 1/2$ and where $a \sim 0.5 $.

We assume the infinite product representation\cite{lav},
\[
{\frac{ Q(\lambda ,m) }{\Gamma(i\lambda )}} = B \; e^{A\, \lambda} \;
\prod_{k=1}^\infty\left(1 - {\frac{\lambda }{\lambda_k}}\right) e^{\frac{%
\lambda }{\lambda_k}} \; .
\]

The asymptotic behaviors (\ref{asiceros}) and (\ref{asintQ}) are consistent
provided (see Appendix B)
\[
A = -i \; \psi(2-m) \; .
\]

In addition, $B = -\pi $ according to eq.(\ref{reprML1}). In summary, we
have,
\begin{equation}  \label{repQpi}
{\frac{ Q(\lambda ,m) }{\Gamma(i\lambda )}} = -\pi \; e^{-i \lambda \,
\psi(2-m)} \; \prod_{k=1}^\infty\left(1 - {\frac{\lambda }{\lambda_k}}%
\right) e^{\frac{\lambda }{\lambda_k}} \; .
\end{equation}
where $\psi(z) $ is the digamma function.

The Bethe Ansatz equations are algebraic equations on  zeroes of the
Baxter function. They follow from the Baxter equation (\ref{ecbpom}) and
this infinite product representation. Eq.(\ref{ecbpom}) can be recasted as,
\[
\left[2\lambda ^2-m(1-m) \right] {\frac{ Q(\lambda ,m) }{\Gamma(i\lambda )}}
= -i( \lambda +i)\; {\frac{ Q(\lambda+i ,m) }{\Gamma(i[\lambda +i])}}
+i\lambda (\lambda -i)^2 \; {\frac{ Q(\lambda-i
,m) }{\Gamma(i[\lambda -i])}}
\]
Setting here $\lambda = \lambda_k $ yields,
\begin{equation}  \label{bae}
{\frac{ \lambda_k \; (\lambda_k -i)^2 }{\lambda_k +i}} = e^{2 \psi(2-m)}
\prod_{l=1}^{\infty}\left[ e^{\frac{2i}{\lambda_l}} {\frac{\lambda_l -
\lambda_k - i }{\lambda_l - \lambda_k + i }}\right] \; .
\end{equation}
We see that the Bethe Ansatz equations are here an infinite system of
algebraic equations. In the present Reggeon problem it is more effective to
solve the Baxter equations by looking at the poles of the Baxter function
[see eq.(\ref{relrec})] rather than to the Bethe Ansatz equations
(\ref{bae}%
). In customary cases the Bethe Ansatz equations are the more effective
tool \cite{rev}.

Since the Baxter function is explicitly known for the Pomeron we can find an
infinite number of sum rules for the zeroes $\lambda_k $ just by matching
eqs. (\ref{f32}) or (\ref{qserie}) with eq.(\ref{repQpi}).

For example, for $\lambda \to i $ we get from eq.(\ref{repQpi}),
\[
-i\lim_{\lambda \rightarrow i}\frac d{d\,\lambda }\ln Q(\lambda ,m)=\frac
i{\lambda -i} + 1 - \gamma + \sum_{k=1}^\infty{\frac{1 }{i \, \lambda_k
\left( 1 + i \, \lambda_k \right)}} - \psi(2-m) \; .
\]
Equating with eq.(\ref{derli}) yields,
\[
\sum_{k=1}^\infty{\frac{1 }{i \, \lambda_k \left( 1 + i \,
\lambda_k\right)}}
= 1 - \gamma - \psi(1-m)-{\frac{1 }{m-1}} \; .
\]

\section{Solving the Baxter equation for the $n$-reggeon state}

The Baxter equation for the odderon takes the form
\bea\label{eqBa3}
&&\left[ 2 \lambda^3 - m(m-1)\lambda  + i \mu \right]
Q\left( \lambda ;\,m,\mu\right)= \cr \cr
&&(\lambda +i)^3 \; Q\left(
\lambda +i ;\, m, \mu\right) +(\lambda -i)^3 \; Q\left( \lambda
-i ;\,m,\mu\right) \; .
\eea
where
\be
q_3 = i \mu \,,\,\, Im \, (\mu )=0 \,.
\ee
The reality property of $\mu$ is needed to obtain  the single-valuedness
of the odderon wave function in the coordinate space \cite{muchos}.

Eq.(\ref{eqBa3}}) can be solved asymptotically for large $ \lambda $
making the power-like ansatz,
\be \label{ansa}
Q\left( \lambda ;\,m,\mu\right)\buildrel{ \lambda \to \infty}\over=
\lambda^a \left[ 1 + {b \over \lambda} + {\cal O}\left({1 \over
\lambda^2}\right) \right]
\ee
Inserting eq.(\ref{ansa}) into eq.(\ref{eqBa3}) yields two solutions:
$ a_+ = m-3 $ and $ a_- = -m-2 $ connected by the $m
\Longleftrightarrow 1-m $ symmetry. The general solution of
eq.(\ref{eqBa3}) will have thus the asymptotic behavior
\bea\label{asiood}
Q\left( \lambda ;\,m,\mu\right)&\buildrel{ \lambda \to \infty}\over=&
A_+ \; \lambda^{m-3} \left[ 1 -i {\mu \over (m-1)(m-2)\lambda} + {\cal
O}\left({1 \over \lambda^2}\right) \right]  + \cr \cr
&+&A_-  \; \lambda^{-m-2} \left[ 1 -i {\mu \over m(m+1)\lambda} + {\cal
O}\left({1 \over \lambda^2}\right) \right] \nonumber
\eea
where $ A_{\pm} $ are constants.

For the pomeron case the Baxter function can be expressed as a sum of
simple poles in the upper plane [see eq.(\ref{reprML1})]. This is not
the case for the odderon. If we try an Ansatz
with only simple poles in the upper plane,
\be\label{qpolsi}
Q_0\left( \lambda \right) = \sum_{r=0}^{\infty} { c_r(m,\mu)
\over
\lambda -r\,i} \,,
\ee
the asymptotic behaviour of this
sum for $\lambda \rightarrow \infty$ turns out to be
\be
Q_0\left( \lambda \right)=c/\lambda \,,\,\, c=\sum_{r=0}^{\infty}
c_r(m,\mu)\,,
\ee
where $c$ is not zero. Therefore $Q_0$ does not
satisfy the Baxter equation for the odderon at $\lambda \rightarrow
\infty$. In particular
the behaviour (\ref{ansa}) is not fulfilled by
$Q_0\left( \lambda \right)$.
Note, that in the case of the Pomeron the simple pole ansatz  indeed
satisfies the Baxter equation at infinity because the sum of residues
vanishes [eq.(\ref{sum0})].

For the odderon, we have to include in the Ansatz double poles. It
will be shown below, that in the general case of $n$ reggeons $Q$
may contain poles of the order $n-1$.
The odderon Baxter function can be written in the form
\be\label{mlodd}
Q\left( \lambda ;\,m,\,\mu\right)=\sum_{r=0}^{\infty} \left[
-{a _r(m,\,\mu) \over (\lambda -r\,i)^2}+i \,{b _r(m,\,\mu) \over
\lambda -r\,i} \right]\,.
\ee
The residues satisfy the recurrence relations
\bea \label{relrodd}
&&a _1(m,\,\mu)=-\mu \,a _0(m,\,\mu)\quad , \quad 8 \,a _2(m,\,\mu)=
\left[ 2+m(m-1)-\mu\right]\,a _1(m,\,\mu) \; , \cr\cr
&&(r+1)^3 \,a _{r+1}(m,\,\mu)=
\left[ 2\,r^3+m(m-1)\,r-\mu\right]\,a _r(m,\,\mu)-(r-1)^3\,
a _{r-1}(m,\,\mu)\; ,     \cr\cr
&&
b _1(m,\,\mu)=-\mu \,b _0(m,\,\mu)+m(m-1)\,a _0(m,\,\mu )-
3\,a_1(m,\,\mu)\,, \cr\cr
&&
8 \,b _2(m,\,\mu)= \left[ 2+m(m-1)-\mu \right]\,b _1(m,\,\mu)
+\left[ 6+m(m-1) \right]\,a _1(m,\,\mu)-12\,a _2(m,\,\mu)\,, \cr\cr
&&
(r+1)^3 \,b _{r+1}(m,\,\mu)= \left[ 2\,r^3+m(m-1)\,r-\mu \right]\,
b _r(m,\,\mu)-(r-1)^3\,b _{r-1}(m,\,\mu) \cr\cr
&&
+\left[ 6\,r^2+m(m-1) \right]\,a _r(m,\,\mu)-
3\,(r+1)^2\,a _{r+1}(m,\,\mu)-3\,(r-1)^2\,a _{r-1}(m,\,\mu)\; .
\eea
We can normalize
\be
a _0(m,\,\mu)=1.
\ee
In order to fulfill the Baxter equation at infinity and to obtain the
asymptotic behaviour (\ref{ansa}), we impose
\be \label{b0}
\sum_{r=0}^{\infty} b _r(m,\,\mu) =0\; .
\ee
This equation fixes the value of $b _0(m,\,\mu)$. Therefore, all
coefficients $ a _n(m,\,\mu) $  and $ b _n(m,\,\mu) $ are univocally
determined.

We find for large $r$ that both $a_r(m,\,\mu)$ and $b_r(m,\,\mu)$
decrease
as $ r^{m-3}$. As in the pomeron case, the asymptotic behaviour of
the
Baxter function for large $ \lambda $ is governed by the late terms in
the Mittag-L\"offler series (\ref{mlodd}). Evaluating the sum of such
late terms we reproduce the asymptotic behaviour (\ref{asiood}).

We see from the recurrence relations (\ref{relrodd}) that
for $\mu \rightarrow 0$ the double poles with $r>0$ disappear
[$a _r(m,\,0)=0$]. Only the double pole at the origin remains and the
solution of the Baxter equation for the
Odderon can be expressed in terms of  that for the Pomeron. With our
normalization we obtain:
\be \label{pomodd}
Q\left( \lambda ;\,m,\,0\right)=-\frac{Q\left( \lambda ,\,m
\right)}{i\,\pi \lambda}\,,
\ee
where $ Q\left( \lambda ,\,m \right) $ is the pomeron Baxter function.

Eq.(\ref{pomodd}) holds because for small $\lambda$
$$
\lim _{\lambda \rightarrow 0}Q\left( \lambda ,\,m \right)={i \pi
\over \lambda}+Q_0(m)\,.
$$
To calculate $Q_0(m)$ one can use eq.(\ref{qcercai}), the relation
\be
Q\left( -i ,\,m \right)=-\pi ^2 \,\frac{m(m-1)}{\sin (\pi m)}
\ee
and the Baxter equation for $Q\left( \lambda ,\,m\right)$ near
$\lambda =0$. Thus, we have
\be
Q_0(m)=\pi \left[\frac{\pi }{\sin (\pi m)}+\psi (m)+\psi (1-m)-2
\psi (1)\right]\,
\ee
in agreement with the asymptotics at $m \rightarrow 2$ [see
eq.(\ref{m2})]
$$
Q\left(\lambda ,\,m \right) \rightarrow {2 \,\pi \over m-2}\,.
$$
Therefore, we obtain the residue $ b_0(m,\,\mu)$ of the simple pole
in the Odderon solution for $\mu =0$ as
\be
b _0(m,\,0)=\frac{\pi }{\sin (\pi
m)}+\psi (m)+\psi (1-m)-2 \psi (1) \,.
\ee
and in particular,
$$
 b_0(\frac12,\,0) = \pi - 4 \, \log 2 = 0.369004\ldots \; .
$$

 From  equation (\ref{b0}) one can compute several
terms of the small-$\mu$ expansion of $b _0(m,\, \mu )$ at $m=1/2$
\be
 b_0(\frac12,\,\mu)= 0.369004-2.835 \,\mu -2.749 \mu ^2 -2.947 \mu ^3+....
\ee

It will be shown in the next sections that the odderon energy
$E$ (related with the intercept $\Delta$ [see eq.(\ref{interc})]) is calculated
in terms of the behaviour of $Q\left(\lambda ;\,m,\,\mu\right)$ and $
Q\left(\lambda^* ;\,\widetilde{m},\,-\mu\right)$
at their singular point $\lambda =i$ and $\lambda^* =i$,
respectively(see (\ref{engen})):
\be \label{Eexa}
E=\frac{b_1(m,\,\mu)}{a_1(m,\,\mu)}+
\frac{b_1(\widetilde{m},\,-\mu)}{a_1(\widetilde{m},\,-\mu)}+6 \,.
\ee
In particular, this gives the possibility to calculate the energy for
$m=1/2$ as a series in $\mu$
\be \label{muchico}
E=b_0(\frac12,\,\mu)+b_0(\frac12,\,-\mu)=0.738008-5.498 \; \mu^2+...
\ee
in agreement with the results by R. Janik and J. Wosiek \cite{YW}
(take into account that we define the energy with an opposite
sign). $E$ is a meromorphic function of $\mu$. For $m=1/2$ its
poles are situated on the real axis at the points
$$
\mu ^{(1)}=\pm 0.91450 \ldots \,,\,\,\mu ^{(2)}=\pm 4.7340 \ldots \,,\,\,\mu
^{(3)}= \pm 13.4 \ldots \,,\,...\,.
$$
The asymptotic behaviour of the energy at large $\mu$ was calculated
in ref. \cite{dual}:
\be \label{mugrande}
\frac{E}{2}=\ln \mu +3\,\gamma +\left[\frac{3}{448}+
\frac{13}{120}\left(m-\frac{1}{2}\right)^2-
\frac{1}{12}\left(m-\frac{1}{2}\right)^4\right]\,\frac{1}{\mu ^2}+\ldots \,.
\ee

We checked that both the small and  large $ \mu $ approximations
given by eqs.(\ref{muchico}) and (\ref{mugrande}), respectively are in
excellent agreement with numerical values obtained from the exact
equations (\ref{relrodd})-(\ref{b0}) and (\ref{Eexa}).

\bigskip

A solution of the Baxter equation independent of eq.(\ref{mlodd})
can be written as follows
\be
Q\left( -\lambda ;\,m,\,-\mu\right)=\sum_{r=0}^{\infty} \left[
-{a _r(m,\,-\mu) \over (\lambda +r\,i)^2}-i \,{b _r(m,\,-\mu) \over
\lambda +r\,i} \right]\,.
\ee
One can verify the relation
$$
\left[Q\left( -\lambda ;\,m,\,-\mu\right)\right]^*=Q\left( \lambda
^* ;\,\widetilde{m},\,-\mu\right)
$$

Furthermore, it turns out that one can construct a solution of the
Baxter equation with simple poles, providing that they are
situated at $\lambda =i\,r$ ($r=0, \,\pm 1, \,\pm 2,...$):
\be
Q_s\left( \lambda ;\,m,\,\mu\right)=\sum_{r=-\infty}^{+\infty}
i\,{c _r(m,\,\mu) \over
\lambda -r\,i} \,.
\ee
We normalise $Q_s$ as follows,
\be
c _0 (m, \, \mu ) =1 \; .
\ee
Then, the residues satisfy the recurrence relations
\be
-\mu =c _1 (m, \, \mu ) -c _{-1} (m, \, \mu )\; ,
\ee
\be
\left[ 2\,r^3+m(m-1)\,r-\mu\right]\,c _r(m,\,\mu)=(r-1)^3\,
c _{r-1}(m,\,\mu)+(r+1)^3\,c _{r+1}(m,\,\mu)\; .
\ee
An additional constraint for $c _{\pm 1}(m,\,\mu)$
is obtained from the condition that in accordance with the Baxter
equation
 $Q_s$
at $\lambda \rightarrow \infty$
should decrease more rapidly than $1/\lambda$:
\be
\sum_{r=-\infty}^{+\infty} c _r(m,\,\mu) =0 \,.
\ee
It is obvious that
$$
Q_s\left( \lambda ;\,m,\,\mu\right)=-Q_s\left( -\lambda ;\,m,\,
-\mu\right)\,,\,\,\left[Q_s\left( \lambda
;\,m,\,\mu\right)\right]^*=-Q_s\left( \lambda ^*;\,m,\,
-\mu\right)\,.
$$

Investigating  the behaviour of the Baxter functions near their poles
we find that the following relation is true,
\be \label{rel}
\left[ i\,\pi \coth{\pi \lambda} +X(m,\,\mu )\right]\,Q_s
\left( \lambda ;\,m,\,\mu\right)=
\frac{c_1(m, \, \mu )}{a_1(m,\,\mu )}\,
Q\left( \lambda ;\,m,\,\mu\right)+
\frac{c_1(m, \, -\mu )}{a_1 (m,\,-\mu )}\,
Q\left( -\lambda ;\,m,\,-\mu\right)\,,
\ee
where $a_1(m,\,\mu )=-\mu $ and
\be \label{X}
X(m,\,\mu )=\frac{b_1(m,\,\mu )}{a_1(m,\,\mu )}-
\frac{d_1(m,\,\mu )}{c_1(m,\,\mu )}=
\frac{d_1(m,\,-\mu )}{c_1(m,\,-\mu )}-
\frac{b_1(m,\,-\mu )}{a_1(m,\,-\mu )} \,.
\ee

The quantities $c_r(\mu ),\,d_r(\mu ),\,e_r(\mu)$ appear in the
expansion of $Q_s\left( \lambda ;\,m,\,\mu\right)$
near the poles at $\lambda =i \,r$
\be \label{cde}
\lim _{\lambda \rightarrow i\,r}
Q_s \left( \lambda ;\,m,\,\mu\right) \rightarrow
i\,\frac{c_r(m,\,\mu )}{\lambda - i \,r}+d_r (m,\,\mu )
-i\,e_r (m,\,\mu)\,(\lambda - i \,r)\,
\ee
and satisfy the following relations
$$
c_{-r}(m,\,-\mu )=c_r(m,\,\mu ),\,
d_{-r}(m,\,-\mu )=-d_r(m,\,\mu ),\,
e_{-r}(m,\,-\mu )=e_r(m,\,\mu )\,.
$$
Due to the property of  holomorphic factorization the Baxter
function in the two-dimensional
$\overrightarrow{\lambda}$-space has the form
$$
Q_{m,\widetilde{m};\,\mu }\left( \overrightarrow{\lambda}
\right)=C_{m,\widetilde{m};\,\mu }^{(s)}\,Q_s \left( \lambda
;\,m,\,\mu\right)\,Q_s\left( \lambda^*;\,\widetilde{m},\,
-\mu \right)+
$$
$$
C_{m,\widetilde{m};\,\mu }^{(1)}\,Q\left( \lambda
;\,m,\,\mu\right)\,
Q\left( \lambda
^*;\,\widetilde{m},\,-\mu\right)+C_{m,\widetilde{m};\,\mu
}^{(2)}\,Q\left( -\lambda ;\,m,\,-\mu\right)\,
Q\left( -\lambda ^*;\,\widetilde{m},\,\mu\right)+
$$
$$
C_{m,\widetilde{m};\,\mu }^{(1s)}\,Q\left( \lambda
;\,m,\,\mu\right)\,
Q_s\left( \lambda
^*;\,\widetilde{m},\,-\mu\right)+C_{m,\widetilde{m};\,\mu
}^{(s2)}\,Q\left( -\lambda ;\,m,\,-\mu\right)\,
Q\left( -\lambda ^*;\,\widetilde{m},\,\mu\right)+
$$
$$
C_{m,\widetilde{m};\,\mu }^{(2s)}\,Q\left( -\lambda
;\,m,\,-\mu\right)\,
Q_s\left( \lambda
^*;\,\widetilde{m},\,-\mu\right)+C_{m,\widetilde{m};\,\mu
}^{(s1)}\,Q_s\left( \lambda ;\,m,\,\mu\right)\,
Q\left( \lambda ^*;\,\widetilde{m},\,-\mu\right)+
$$
\be \label{bibax}
C_{m,\widetilde{m};\,\mu }^{(12)}\,Q\left( \lambda
;\,m,\,\mu\right)\,
Q\left( -\lambda
^*;\,\widetilde{m},\,\mu\right)+C_{m,\widetilde{m};\,\mu
}^{(21)}\,Q\left( -\lambda ;\,m,\,-\mu\right)\,
Q\left( \lambda ^*;\,\widetilde{m},\,-\mu\right)\,,
\ee
where we took into account that $q_3^*=-i\mu$.

The coefficients
$C^{(k)}$  are fixed by the condition
of the normalizability of $Q_{m,\widetilde{m};\mu }\left(
\overrightarrow{\lambda}
\right)$, which reduces to the requirement
for $Q_{m,\widetilde{m};\,\mu }
\left( \overrightarrow{\lambda}
\right)$ to be regular at $\sigma =0$ provided that
$\lambda = \sigma +i \,N/2$ with $|N|>0$. For $N=0$ the
poles at $\sigma =0$ are killed by the corresponding factor
in the integration measure.

It is obvious, that
$$
C_{m,\widetilde{m};\,\mu }^{(12)}=
C_{m,\widetilde{m};\,\mu }^{(21)}=0 \,,
$$
because in the opposite case one can not  cancel the
fourth order poles in the product of the corresponding
holomorphic and anti-holomorphic functions.

Further, the following equality
$$
C_{m,\widetilde{m};\,\mu }^{(1)}=
-C_{m,\widetilde{m};\,\mu }^{(2)} \,.
$$
is valid.
To show it,
let us investigate  the Baxter function $Q\left(
\lambda ;\,m,\,\mu\right)$ near the regular points $\lambda =-i\,r$
($r=1,2,...$):
\be
\lim _{\lambda \rightarrow -i\,r} Q\left( \lambda ;\,m,\,\mu\right)=
A_r\left(m,\,\mu\right)+i(\lambda +ir)\,B_r\left(m,\,\mu\right)\,.
\ee

It can be verified, that $A_r\left(m,\,\mu\right)$ and
$B_r\left(m,\,\mu\right)$ for $r>2$ satisfy the same recurrence
relations as
$a_r\left(m,\,-\mu\right)$ and $b_r\left(m,\,-\mu\right)$
respectively. Therefore $A_r(m,\,\mu )$ should be proportional to
$a_r(m,\,-\mu )$
(for $r>0$)
\be
A_r\left(m,\,\mu\right)=\alpha
\left(m,\,\mu\right)\, a_r\left(m,\,-\mu\right)\,.
\ee
But $B_r(m,\,\mu )$ are not proportional to $b_r(m,\,-\mu )$ even if
we would chose
$\mu$ in such way, that
$B_1\left(m,\,\mu\right)= \alpha
\left(m,\,\mu\right)\,b_1\left(m,\,-\mu\right)$.
The reason is that according to the Baxter equation the coefficient
$B_2\left(m,\,\mu\right)$ is expressed not only in
terms of $A_1\left(m,\,\mu\right)$, $A_2\left(m,\,\mu\right)$
and $B_1\left(m,\,\mu\right)$ (similar to $b_2\left(m,\,\mu\right)$),
but it contains also a contribution proportional to
$a_0\left(m,\,\mu\right)=1$ from the pole $1/\lambda ^2$. Therefore
$B_r\left(m,\,\mu\right)$ for $r>1$ are not proportional to
$b_r\left(m,\,-\mu\right)$. From the Baxter equation we can
obtain the following relations
$$
B_r\left(m,\,\mu\right)=\alpha
\left(m,\,\mu\right)\, b_r\left(m,\,-\mu\right)+
\left[B_1\left(m,\,\mu\right)-\alpha
\left(m,\,\mu\right)\, b_1\left(m,\,-\mu\right)\right]\,\frac{
a_r\left(m,\,-\mu\right)}{ a_1\left(m,\,-\mu\right)}+\widetilde{
a}_r\left(m,\,-\mu\right) \,,
$$
where $\widetilde{a}_r\left(m,\,-\mu\right)$ satisfies the same
recurrent relations as $a_r\left(m,\,-\mu\right)$ for $r>1$
with different initial conditions:
$$
\widetilde{a}_1\left(m,\,-\mu\right)=0\,,\,\,\widetilde{
a}_2\left(m,\,-\mu\right)=\frac{1}{8}\,.
$$
Because in the other bilinear contributions to eq. (\ref{bibax})
the residues of the poles in $\sigma$ do not contain $\widetilde{
a}_r\left(m,\,-\mu\right)$, we should cancel them in the
following combination
$$
\lim_{\lambda \rightarrow i \, r}\,\left[Q\left( \lambda
;\,m,\,\mu\right)\,Q\left( \lambda
^*;\,\widetilde{m},\,-\mu\right)-Q\left( -\lambda
;\,m,\,-\mu\right)\,
Q\left( -\lambda ^*;\,\widetilde{m},\,\mu\right)
\right]=
$$
$$
-\frac{1}{\sigma ^2}\,\left[\alpha
\left(\widetilde{m},\,-\mu\right)-\alpha
\left(m,\,-\mu\right)\right]\,a_r\left(m,\,\mu\right)\,
a_r\left(\widetilde{m},\,\mu\right)+
\frac{i}{\sigma} \,D_r(m,\,\widetilde{m},\,\mu)\,,
$$
where
$$
D_r(m,\,\widetilde{m},\,\mu )=\left[\alpha
\left(\widetilde{m},\,-\mu\right)+\alpha
\left(m,\,-\mu\right)\right]\,
\left[b_r\left(m,\,\mu\right)\,
a_r\left(\widetilde{m},\,\mu\right)-a_r\left(m,\,\mu\right)\,
b_r\left(\widetilde{m},\,\mu\right)\right]+
$$
$$
\left(\frac{B_1\left(m,\,-\mu\right)-\alpha (m,\,-\mu )\,
b_1(m,\,\mu
)}{a_1\left(m,\,\mu\right)}-
\frac{B_1\left(\widetilde{m},\,-\mu\right)-\alpha
(\widetilde{m},\,-\mu )\,
b_1(\widetilde{m},\,\mu
)}{a_1\left(\widetilde{m},\,\mu\right)}\right)
a_r\left(m,\,\mu\right)\,
a_r\left(\widetilde{m},\,\mu\right)
$$
$$
+a_r\left(\widetilde{m},\,\mu\right)\,
\widetilde{a}_r\left(m,\,\mu\right)-
\widetilde{a}_r\left(\widetilde{m},\,\mu\right)\,
a_r\left(m,\,\mu\right) \,.
$$

According to the relations
$$
a_r(\widetilde{m}, \, \mu )=\left[a_r(m, \, \mu )\right]^*\,,\,\,
\widetilde{a}_r(\widetilde{m}, \, \mu )=\left[\widetilde{a}_r(m, \,
\mu )\right]^*\,,
$$
the contribution containing $\widetilde{a}$ is pure imaginary and
anti-symmetric to the transmutation $m \leftrightarrow
\widetilde{m}$.

In the case of conformal spin $n=m-\widetilde{m}=0$ the function
$$
Q_{m,\,m;\,\mu }\left( \overrightarrow{\lambda}
\right)=Q\left( \lambda
;\,m,\,\mu\right)\,Q\left( \lambda
^*;\,m,\,-\mu\right)-Q\left( -\lambda
;\,m,\,-\mu\right)\,
Q\left( -\lambda ^*;\,m,\,\mu\right)
$$
does not contain poles at $\sigma =0$ for $|N|>0$ and can be
normalized. In the general case $m \neq \widetilde{m}$ to cancel the first and
second order poles at $\sigma =0$ one should take into account
all contributions in eq. (\ref{bibax}).

Let us attemp to construct a normalized wave function for $m \neq
\widetilde{m}$ including all contributions in eq. (\ref{bibax}) except
the second and third terms and the last two terms. That is, we impose,
$$
C_{m,\widetilde{m};\,\mu }^{(1)}=
C_{m,\widetilde{m};\,\mu }^{(2)}=0 \,.
$$
We call such wave function $\Delta Q_{m,\widetilde{m};\,\mu }\left(
\overrightarrow{\lambda} \right)$.

Using  the Baxter equation one can obtain the recurrence relations
for the coefficients $c_r,\,d_r,\,e_r$ of the Laurent expansion
(\ref{cde}) of
$Q_s\left( \lambda ;\,m,\,\mu\right)$ near the pole $\lambda =i\,r$.
They are similar to the relations for the
expansion coefficients $a_r, \,b_r$ and $E_r$ for
$Q\left( \lambda ;\,m,\,\mu\right)$:
\be
\lim _{\lambda \rightarrow i\,r}
Q \left( \lambda ;\,m,\,\mu\right) \rightarrow
-\frac{a_r(m,\,\mu )}{(\lambda - i \,r)^2}+i\,
\frac{b_r (m,\,\mu )}{\lambda - i \,r}
+\,E_r (m,\,\mu) \,.
\ee
We obtain the following relations
$$
c_r(m,\,\mu )=\frac{c_1(m,\,\mu )}{a_1(m,\,\mu )}\,a_r(m,\,\mu )\,,
$$
$$
d_r(m,\,\mu )=
\frac{c_1(m,\,\mu )}{a_1(m,\,\mu )}\,b_r(m,\,\mu )+
\frac{d_1(m,\,\mu )-\frac{b_1(m,\,\mu )}{a_1(m,\,\mu )}\,
c_1(m,\,\mu )}{a_1(m,\,\mu )}\,a_r(m,\,\mu )\,,
$$
$$
e_r(m,\,\mu )=\frac{c_1(m,\,\mu )}{a_1(m,\,\mu )}\,E_r(m,\,\mu )+
\frac{d_1(m,\,\mu )-\frac{b_1(m,\,\mu )}{a_1(m,\,\mu )}\,
c_1(m,\,\mu )}{a_1(m,\,\mu )}\,b_r(m,\,\mu )+
$$
$$
\frac{e_1(m,\,\mu )-\frac{c_1(m,\,\mu )}{a_1(m,\,\mu )}\,E_1(m,\,\mu )-
\left( d_1(m,\,\mu )-\frac{b_1(m,\,\mu )}{a_1(m,\,\mu )}\,
c_1(m,\,\mu )\right)\,
\frac{b_1(m,\,\mu )}{a_1(m,\,\mu )}}{a_1(m,\,\mu )}\,a_r(m,\,\mu )\,.
$$
These relations allow one to verify, that the
coefficients $C_{m,\widetilde{m};\,\mu }^{(t)}$ in the above expression
(\ref{bibax}) can be chosen in such way to cancel all poles at
$\sigma =0$ for $|N|>0$, which leads to the following expression
for $\Delta Q_{m,\widetilde{m};\,\mu }\left(
\overrightarrow{\lambda}
\right)$
$$
\Delta Q_{m,\widetilde{m};\,\mu }\left( \overrightarrow{\lambda }
\right)=-\left[X(m,\,\mu)+X(\widetilde{m},\,\mu) \right]\,
Q_s \left( \lambda
;\,m,\,\mu \right)\,Q_s\left( \lambda^*;\,\widetilde{m},
\,-\mu \right)+
$$
$$
\frac{c_1(m,\,\mu )}{a_1(m,\,\mu )}\,Q \left( \lambda
;\,m,\,\mu\right)\,Q_s\left( \lambda^*;\,\widetilde{m},\,
-\mu \right)-\frac{c_1(\widetilde{m},\,\mu )}{a_1(
\widetilde{m},\,\mu )}\,
Q_s \left( \lambda
;\,m,\,\mu \right)\,Q\left( -\lambda^*;\,\widetilde{m},\,
\mu \right)+
$$
\be
\frac{c_1(m,\,-\mu )}{a_1(m,\,
-\mu )}\,Q \left(-\lambda
;\,m,\,-\mu\right)\,Q_s\left( \lambda^*;\,\widetilde{m},\,
-\mu \right)-\frac{c_1(\widetilde{m},\,-\mu )}{a_1(
\widetilde{m},\,-\mu )}\,Q_s \left( \lambda
;\,m,\,\mu \right)\,Q\left( \lambda^*;\,\widetilde{m},\,
-\mu \right) \,,
\ee
where $X(m,\,\mu)$ is defined in eq. (\ref{X}).

Note, that the expression $\Delta Q_{m,\widetilde{m};\,\mu }\left(
\overrightarrow{\lambda } \right)$ constructed above is in fact
zero due to  eq.(\ref{rel}). However, the wave function
$Q_{m,\widetilde{m};\,\mu }\left(\overrightarrow{\lambda } \right)$
given by eq.(\ref{bibax}) is normalizable and does not vanish when
  all contributions are included. That is, choosing
$ C_{m,\widetilde{m};\,\mu }^{(1)}\neq 0 \neq C_{m,\widetilde{m};\,\mu
}^{(2)} $ (but excluding the last two terms). Vanishing $\Delta
Q_{m,\widetilde{m};\,\mu }\left(
\overrightarrow{\lambda } \right)$ allows us diminish the number
of independent bilinear combinations of the Baxter functions.

\bigskip

It is important, that we constructed the  normalized function
$Q_{m,\widetilde{m};\,\mu }\left( \overrightarrow{\lambda }
\right)$ without imposing any condition on the numerical
value of $\mu$. This function is a bilinear combination
of  different Baxter functions in the holomorphic and
anti-holomorphic spaces. Let us take into account the
physical requirement, that all these Baxter functions have
the same energy, because in the opposite case
$Q_{m,\widetilde{m};\,\mu }\left( \overrightarrow{\lambda }
\right)$ would not have a definite total energy. According to the
results of the next sections the
energy is expressed in terms of the sum of logarithmic derivatives
of the functions $(\lambda -i)^2Q(\lambda)$ at $\lambda =i$
in the holomorphic and anti-holomorphic spaces. We have two
independent functions with second order poles at $\sigma =i$.
They are
$Q \left( \lambda
;\,m,\,\mu\right)$ and $\coth (\pi \lambda) \,Q_s
\left( \lambda ;\,m,\,\mu\right)$.
The equality of the energies calculated from these
functions gives the quantization condition for $\mu$:
\be
-2 \, X(m,\,\mu )=
\frac{d_1(m,\,\mu )}{c_1(m,\,\mu )}-
\frac{d_1(m,\,-\mu )}{c_1(m,\,-\mu )}-
\frac{b_1(m,\,\mu )}{a_1(m,\,\mu )}
+\frac{b_1(m,\,-\mu )}{a_1(m,\,-\mu )} =0\,.
\ee
We found from the above equations numerically the first roots for
$m=
\widetilde{m}=1/2$:
$$
\mu _1 = 0.205257506 \ldots \quad , \quad \mu _2= 2.3439211 \ldots
\quad , \quad\mu _3=8.32635\ldots
\quad , \quad\mu _4=20.080497\ldots \quad ,\ldots
$$
with the corresponding energies
$$
E_1=0.49434 \ldots \quad , \quad
E_2=5.16930 \ldots \quad , \quad
E_3 = 7.70234 \ldots \quad , \quad
E_4 = 9.46283  \ldots \quad ,  \ldots
$$
These values are in a full agreement with the results of R. Janik,
J. Wosiek and other authors (see \cite{muchos}) obtained by the
diagonalization
of the integral of motion $q_3$ in the impact parameter space and
imposing the property of the single-valuedness of the wave function.

\bigskip

Let us now consider the Baxter equation for the $n$-reggeon
composite state:
\be
\Lambda ^{(n)}(\lambda ;\;\vec{\mu})\,
Q\left( \lambda ;\,m,\vec{\mu}\right)=
(\lambda +i)^n \; Q\left(
\lambda +i ;\,m, \vec{\mu}\right) +(\lambda -i)^n \; Q\left(
\lambda
-i ;\,m,\vec{\mu}\right)
\; ,
\end{equation}
where $\Lambda ^{(n)}(\lambda )$ is the polynomial
\be
\Lambda ^{(n)}(\lambda ;\;\vec{\mu})=\sum _{k=0}^n  (-i)^k\,\mu
_k\,\lambda
^{n-k}\,,\;\mu _0=2,\; \mu _1=0,\;\mu _2=m(m-1)\,,
\ee
where we assume, that $\mu _k=i^k\,q_k$ for $k>2$ are real
numbers. The last condition is needed
in order to have
a normalizable  wave functions.

We search the solution of this equation in the form of a sum over
the poles
of the orders from $1$ up to $n-1$:
\be
Q\left( \lambda ;\,m,\vec{\mu}\right)=\sum _{r=0}^\infty
\frac{P^{(n-2)}_{r;m,\vec{\mu}}(\lambda )}{(\lambda
-i\,r)^{n-1}}\,.
\ee

Putting this ansatz in the equation, we obtain recurrence relations
for polynomials $P^{(n-2)}_{r;m,\vec{\mu}}(\lambda )$ of the
order $n-2$ allowing to
calculate them providing that $P^{(n-2)}_{0;m,\vec{\mu}}(\lambda )$
is known.
Indeed, let us define  the expansion of a function
$f(\lambda )$ in the
power series over $\lambda -i\,r$ up to the order $n-2$:
\be
\left(f(\lambda )\right)^{(n-2)}_r=(\lambda
-i\,r)^{n-2}\lim
_{\lambda \rightarrow i\,r}\frac{f(\lambda )}{(\lambda
-i\,r)^{n-2}}.
\ee
Then the recurrence relations for the coefficients of polynomials
$P^{(n-2)}_{r;m,\vec{\mu }}(\lambda )$ can be written as follows
$$
\left(\Lambda ^{(n)}(\lambda ,\; \vec{\mu
})P^{(n-2)}_{r;m,\vec{\mu
}}(\lambda )\right)^{(n-2)}_r=
$$
\be
\left((\lambda +i)^n
P^{(n-2)}_{r+1;m,\vec{\mu }}(\lambda
+i)\right)_r^{(n-2)}+\left((\lambda
-i)^n
P^{(n-2)}_{r-1;m,\vec{\mu }}(\lambda -i)\right)_r^{(n-2)} \,.
\ee

We can normalize the solution imposing the constraint
\be
\lim _{\lambda \rightarrow i}P^{(n-2)}_{0;m,\vec{\mu
}}(\lambda )=1
\ee
Then the other independent coefficients of the polynomial
$P^{(n-2)}_{0,m,\vec{\mu }}(\lambda )$ are determined from the
condition
\be
\lim _{\lambda \rightarrow \infty} Q\left( \lambda
;\,m,\vec{\mu}\right) \sim \lambda ^{-n+m}\,,
\ee
necessary  to provide $ Q\left( \lambda
;\,m,\vec{\mu}\right)$ to be a solution of the Baxter equation at
$\lambda \rightarrow \infty$. According to the Baxter equation this
condition is
fulfilled if
\be
\lim_{\lambda \rightarrow \infty} \lambda ^{n-2} \,\sum
_{r=0}^\infty
{P^{(n-2)}_{r;m,\vec{\mu}}(\lambda ) \over (\lambda -i\,r)^{n-1}}
=0 \,.
\ee
It gives $n-2$ linear equations giving a possibility to calculate
all coefficients of the polynomial $P^{(n-2)}_{0,m,\vec{\mu
}}(\lambda
)$.

The existence of the other independent solution
\be
Q\left( -\lambda ;\,m,\vec{\mu^s}\right)=\sum _{r=0}^\infty
\frac{P^{(n-2)}_{r;m,\vec{\mu ^s}}(\lambda )}{(-\lambda
-i\,r)^{n-1}}\,,
\ee
where $\vec{\mu^s}$ has the components $\mu ^s _k=(-1)^k \mu _k$,
is related with the symmetry of the Baxter equation to the
simultaneous transformations
$$
\lambda \rightarrow -\lambda \,,\,\,\mu \rightarrow \mu ^s \,.
$$
One can verify, that
$$
\left( Q\left( -\lambda ;\,m,\vec{\mu^s}\right)\right)^*=Q\left(
\lambda ^*;\,\widetilde{m},\vec{\mu^s}\right)\,.
$$

Let us investigate now the behaviour of the Baxter function near
the regular points $\lambda =-i\,r$ ($r=1,2,...$):
\be
\lim _{\lambda \rightarrow -i\,r}\frac{ Q\left( \lambda
;\,m,\,\vec{\mu}\right)}{(\lambda +i\,r)^{n-2}}=\frac{
S^{(n-2)}_ {r;m,\vec{\mu}}(\lambda )}{(\lambda +i\,r)^{n-2}}\,,
\ee
where $S^{(n-2)}_ {r;m,\vec{\mu}}(\lambda)$ are polynomials obeying
certain recurrence relations which can be obtained from the Baxter
equation. These
reccurence relations for $r> 2$ are the same as for $P^{(n-2)}_
{r;m,\vec{\mu ^s}}(-\lambda)$, but we can not chose these
two functions
to be proportional even by imposing this proportionality at $r=1$
by an appropriate choice of the integrals of motion $\mu _k$. Similar
to the case of the odderon it is related with the fact, that
$S^{(n-2)}_ {2;m,\vec{\mu}}(\lambda)$ contains in these recurrence
relations an additional contribution from the pole $\lambda ^{1-n}$.
Therefore to cancel the corresponding singularities
in the wave function $Q_{m,\,\widetilde{m},
\,\vec{\mu}}\left( \overrightarrow{\lambda}\right)$ at $\sigma =0$
the bilinear
combinations of the above functions $Q_{m,\,\vec{\mu}}(\lambda )$
and $Q_{m,\,\vec{\mu^s}}(-\lambda )$ should be in the form
$$
Q\left( \lambda ;\,m,\vec{\mu }\right)\,Q\left( \lambda ^*
;\,\widetilde{m},\vec{\mu^s}\right)-Q\left( -\lambda ;\,m,\vec{\mu
^s}\right)\,Q\left( -\lambda ^*;\,\widetilde{m},\vec{\mu }\right)
\,.
$$
To cancel other pole singularities we
should introduce a set of
additional Baxter function having the poles simultaneosly in the
upper
and low semi-planes of the complex $\lambda$-plane.
$$
Q^{(t)}\left( \lambda ;\,m,\vec{\mu}\right)=\sum _{r=0}^\infty
\left( \frac{P^{(t-1)}_{r;m,\vec{\mu}}(\lambda )}{(\lambda
-i\,r)^t}+\frac{P^{(n-2-t)}_{r;m,\vec{\mu}}(-\lambda )}{(-\lambda
-i\,r)^{n-1-t}}\right)\,,
$$
where the polynomials $P^{(t-1)}$ and $P^{(n-2-t)}$
are fixed by the reccurence relations following
from the Baxter equation and by the condition, that the new Baxter
functions decrease at infinity  more rapidly than
$\lambda ^{-n+2}$. These functions are linear combinations of
$Q\left( \lambda ;\,m,\vec{\mu }\right)$ and $Q\left( -\lambda
;\,m,\vec{\mu^s}\right)$
with the coefficients depending on
$\coth (\pi \lambda)$. Using all these functions in the
holomorphic and anti-holomorphic spaces one can construct
$Q_{m,\,\widetilde{m},\,\vec{\mu}}
\left( \overrightarrow{\lambda}\right)$ without the
poles at $\sigma =0$. The quantization condition for $\mu$ is
obtained from the requirement, that the energy should be the same
for all Baxter functions $Q^{(t)}\left( \lambda
;\,m,\vec{\mu}\right)$.
We calculate the spectrum of the reggeon states for $n>3$ in a
forthcoming paper.

\section{Hamiltonian in the BS representation}

The high energy asymptotics of the scattering amplitude corresponding to the
contribution related to the $t$-channel exchange of the composite state of
$%
n $ reggeized gluons in the multi-color QCD has the form

\be \label{interc}
A(s,t)\sim i^{n-1\,}s\,s^\Delta \,,\,\,\Delta =-\frac{g^2}{8\pi
^2}N_c\,E\,,
\ee
where $E$ is the ground state energy for the Schr\"odinger equation

\[
E\,\Psi _{m,\widetilde{m}}(\overrightarrow{\rho
_1},\,\overrightarrow{\rho
_2%
},\,\ldots,\overrightarrow{\rho _n};\overrightarrow{\rho _0})=
H\,\Psi _{m,%
\widetilde{m}}(\overrightarrow{\rho _1},
\,\overrightarrow{\rho _2},\,\ldots,%
\overrightarrow{\rho _n};\overrightarrow{\rho _0})\,.
\]

The Reggeon Hamiltonian is

\[
H=\frac 12\sum_{k=1}^n\,H_{k,k+1}\,.
\]
Here $1/2$ is the ration of the color factors for the adjoint and singlet
representations of the color group and the pair BFKL Hamiltonian is given
by
\begin{eqnarray}
H_{1,2}&=&\ln \left| p_1\right| ^2+\ln\left| p_2\right| ^2+ \cr \cr &+&
\frac{p_1p_2^{*}}{\left| p_1\right| ^2\left| p_2\right| ^2} \; \ln \,\left|
\rho _{12}\right| ^2p_1^{*}p_2+\frac{p_2p_1^{*}}{\left| p_1\right| ^2\left|
p_2\right| ^2}\; \ln\left| \rho _{12}\right|^2 \, p_2^{*}p_1-4 \, \psi
(1)\,.
\end{eqnarray}
It enjoys the property of holomorphic separability
\[
H_{1,2}=h_{12}+h_{12}^{*}\,,
\]
where
\[
h_{12}=\ln (p_1p_2)+\frac 1{p_1}\; (\ln\rho_{12}) \; p_1+\frac 1{p_2}\;
(\ln\rho_{12}) \; p_2-2 \, \psi (1)\,.
\]
We now perform the unitary transformation of the hamiltonian to the BS
representation, where $P$ and the roots $\widehat{\lambda }_1,\,\widehat{%
\lambda }_2,\ldots,\widehat{\lambda }_{n-1}$ of the equation $B(u)=0$ are
diagonal operators. In this new representation both the integrals of motion
and the hamiltonian should have simple separability properties.

Let us start with the case of the Pomeron, where

\[
H=H_{1,2}\,.
\]
We now calculate the action of the hermitially conjugated
Hamiltonian on the
eigenfunctions of the operator $B(u)$ given by eq.(\ref{baseN})
(with $\lambda =-\lambda _1=\sigma +iN/2$,
\begin{eqnarray}
&&H_{1,2}^{+}\,\left( \frac p{1-p}\right) ^{-i\lambda ^
{*}}\left( \frac{p^{*}
}{1-p^{*}}\right) ^{-i\lambda }=\ln \left[ \frac{\left| p\right| ^2\left|
1-p\right| ^2}{m^4}\right] \,\left( \frac p{1-p}\right) ^{-i\lambda
^{*}}\left( \frac{p^{*}}{1-p^{*}}\right) ^{-i\lambda } - \cr \cr &&-\frac
1\pi \int d^2k \; \frac{\left[ k \, p^{*}(1-k^{*})(1-p)+k^{*} \, p \,
(1-k)(1-p^{*})\right]} {\left( \left| k-p\right| ^2+ m^2 \right)
\,\left| k\right| ^2\left| 1-k\right| ^2}\left( \frac k{1-k}\right)
^{-i\lambda ^{*}}\left( \frac{k^{*}}{1-k^{*}}\right) ^{-i\lambda }\,,
\nonumber
\end{eqnarray}
where $P = 1$  and $m \rightarrow 0$ is an
infrared regulator, corresponding to the vector boson mass rescaled by $P$ (cf.
\cite{BFKL}).

Using the anti-Wick rotation of momenta $k_2=-ik_0$ and $p_2=-ip_0$
as in eq.(%
\ref{antiW})  after some transformations we can write the result in
holomorphically separable form
\begin{eqnarray}
&&H_{1,2}^+\,\left( \frac p{1-p}\right) ^{-i\lambda ^{*}}\left( \frac{p^{*}}{
1-p^{*}}\right) ^{-i\lambda }\cr \cr
&&= \left( \frac{p^{*}}{1-p^{*}}\right)
^{-i\lambda }h_{12}\,\left( \frac p{1-p}\right) ^{-i\lambda ^{*}}+\left( \frac
p{1-p}\right)^{-i\lambda ^{*}}h_{12}^{*}\left( \frac{p^{*}}{1-p^{*}}\right)
^{-i\lambda }\; , \nonumber
\end{eqnarray}
where
\begin{eqnarray}
&&h_{12}\,\left( \frac p{1-p}\right) ^{-i\lambda ^{*}}\cr \cr
&&= \left[ \ln \,\frac{p(1-p)}{\varepsilon ^2}+\pi \,i\,\coth (\pi
\lambda ^{*})\right]
\left( \frac p{1-p}\right) ^{-i\lambda ^{*}}-\int_{p+\varepsilon }^1 dk \;
\frac{\left( p+k-2\,k\,p\right) \,k^{-1-i\lambda ^*}}{ (k-p)\,(1-k)^{1-i%
\lambda ^{*}}}\nonumber
\end{eqnarray}
and
\begin{eqnarray}
&&h_{12}^{*}\,\left( \frac{p^{*}}{1-p^{*}}\right) ^{-i\lambda }\cr \cr
&&=\left[ \ln \,\frac{p^{*}(1-p^{*})}{\varepsilon ^2}+\pi \,i\,\coth (\pi
\lambda )\right] \left( \frac{p^{*}}{1-p^{*}}\right) ^{-i\lambda
}-\int_{p^{*}+\varepsilon }^1d\,k^{*} \; \frac{\left(
p^{*}+k^{*}-2\,k^{*}\,p^{*}\right) \,(k^*)^{-1-i\lambda}}{
(k^{*}-p^{*})\,(1-k^{*})^{1-i\lambda ^{*}}} \,. \nonumber
\end{eqnarray}
Here $\varepsilon \rightarrow 0$ is an intermediate infrared cut-off. The
hamiltonians $h_{12}$ and $h_{12}^{*}$ have branch point singularities at $%
p=0,\,1,\,\infty $ and $p^{*}=0,\,1,\,\infty $, respectively, but the total
hamiltonian $H_{1,2}$ is single-valued.

We obtain for $\left| p\right| \rightarrow 0$:
\begin{eqnarray}
&&\lim_{\left| p\right| \rightarrow 0}H_{1,2}\,\left( p\right) ^{-i\lambda
^{*}}\left( p^{*}\right) ^{-i\lambda }\cr \cr
&&=\left[ -\ln \left| p\right|
^2+\psi (1+i\lambda ^{*})+\psi (1-i\lambda ^{*})+\psi (1+i\lambda )+\psi
(1-i\lambda )-4\psi (1)\right] \,(p)^{-i\lambda ^{*}}(p^*)^{-i\lambda
} \; .\nonumber
\end{eqnarray}
Taking into account that in the integral
\[
\Psi _{m,\widetilde{m}}(\overrightarrow{p},\,\overrightarrow{1}-
\overrightarrow{p})=\frac 1{2\pi ^2}\int_{-\infty }^{+\infty} d\,\sigma
\sum_{N=-\infty }^{+\infty} \left( \frac p{1-p}\right)
^{-i\lambda ^*}\left( \frac{
p^{*}}{1-p^{*}}\right) ^{-i\lambda }\,\Phi
_{m,\widetilde{m}}(
\overrightarrow{1},\,\overrightarrow{\lambda })\,,\,\,\,\lambda = \sigma
+i \frac{N}{2}
\]
for $\left| p\right| \rightarrow 0$ the leading asymptotics corresponds to $%
N=0$, and shifting the integration contour in $\sigma $ in the upper
half-plane up to the first singularity of $\Psi _{m,\widetilde{m}}(%
\overrightarrow{1},\,\overrightarrow{\lambda })$, corresponding to the poles
at $\lambda ,\lambda ^{*}=i$, we obtain for the hamiltonian near these
singularities
\[
\lim_{\lambda ,\lambda ^{*}\rightarrow i}H_{1,2}\, \Phi_{m,\widetilde{m}}(
\overrightarrow{1},\,\overrightarrow{\lambda })=\lim_{\lambda ,\lambda
^{*}\rightarrow i} \left[ i\left( \frac
\partial {\partial \lambda }+\frac \partial {\partial \lambda ^{*}}+\frac
1{\lambda -i}+\frac 1{\lambda ^{*}-i}\right) +2\right] \Phi_{m,\widetilde{m}
}(\overrightarrow{1},\,\overrightarrow{\lambda })\,.
\]
Using eq.(\ref{fiqq}) for the Pomeron wave function in this limit,
\[
\Phi _{m,\widetilde{m}}(\overrightarrow{1},\,\overrightarrow{\lambda })\sim
Q(\lambda ,m)\,Q(\lambda ^{*},\widetilde{m})\,\left| \lambda \right|
^2 \; ,
\]
the Pomeron energy is given as follows
\[
E_{12}=i\lim_{\lambda ,\lambda ^{*}\rightarrow i}\left\{ \frac \partial
{\partial \lambda }\ln \left[ (\lambda -i) \lambda^2 \, Q(\lambda
,m)\right] +\frac \partial {\partial \lambda^{*}}\ln \left[  (\lambda^*
-i) (\lambda^*)^2 \,Q(\lambda^{*},\widetilde{m})\right] \right\} \,.
\]
We obtain for the Pomeron energy using the behavior for $Q(\lambda ,m)$
near $\lambda = i $ [eq.(\ref{qcercai})]
\be \label{enpom}
E_{12}=\psi (m)+\psi (1-m)+\psi (\widetilde{m})+\psi (1-\widetilde{m})-4\psi
(1)
\ee
in agreement with the known result \cite{BFKL}.

\section{Energy for multi-reggeon composite states}

Let us investigate the behavior of the wave functions for the composite
states in the region where the values of gluon momenta are strictly ordered:
\[
|p_1|<<|p_2|<<\,\ldots\, <<|p_n| \,.
\]

To begin with, we consider the Odderon case, where the wave function is given
by
\[
\Psi _{m,\widetilde{m}}(\overrightarrow{p_1},\,\overrightarrow{p_2},\,%
\overrightarrow{p_3})=
\]
\[
\prod_{r=1}^3 \left( \overrightarrow{p_r} \right)^2 \int \prod_{k=1}^3
\left[ \frac{d^2\rho _k}{2\pi } \, \exp (i\overrightarrow{p_k}\cdot%
\overrightarrow{\rho_{k0}}) \right] \, \left( \frac{\rho _{23}}{\rho
_{20}\rho _{30}}\right)^m \left(\frac{\rho_{23}^{*}}{\rho _{20}^{*}\rho
_{30}^{*}}\right)^{\widetilde{m}} \phi_{m,\widetilde{m}}(x\,,x^{*}),
\]
where

\[
x=\frac{\rho _{12}\rho _{30}}{
\rho _{10}\rho _{32}}\,
\]
and the function $\phi _{m,\widetilde{m}}(x\,,x^{*})$ has the property of
the holomorphic factorization

\[
\phi _{m,\widetilde{m}}(x\,,x^{*})=\sum_{i,k}c_{ik}\,\phi _m^i(x\,)\,\phi
_{%
\widetilde{m}}^k(x^{*})\,+(q_3\leftrightarrow -q_3).
\]

The functions $\phi _m^i(x),\,\phi _{\widetilde{m}}^k(x^{*})$ are
independent eigenfunctions of the integral of motion \cite{lip1}
\[
A_m\phi _m^i(x\,)=a_{1-m}\,a_m\,\phi _m^i(x\,)\,=q_3\phi _m^i(x\,)\,,
\]
where $A_m$ is given by
\[
A_m=i^3\,x(1-x)\,\left[ x(1-x)\,\partial ^2+(2-m)\,(1-2x)\,\partial
-(2-m)\,(1-m)\right] \,\partial \,.
\]
The operators
\[
a_m=x\,(1-x)\,(i\partial )^{m+1}\,,\,\,a_{1-m}=x\,(1-x)\,(i\partial
)^{2-m}
\]
perform the duality transformation \cite{dual}.

The three independent eigenfunctions $\phi _m^i(x)$ have the following
small-$x$ asymptotics \cite{muchos}:
\[
\phi _m^1(x\,)\simeq x+O(x^2) \quad , \quad \phi _m^2(x\,)\simeq
1+O(x\ln x) \quad ,\quad  \phi_m^3(x\,)\simeq x^m\left[ 1+O(x)\right] \; ,
\]
which correspond to the following asymptotics of $\phi _{m,\widetilde{m}
}(x\,,x^{*})$ enjoying single-valuedness at the singular points,
\[
\lim_{x\rightarrow 0}\phi _{m,\widetilde{m}}(x\,,x^{*})\simeq x^m \; x^{*
\widehat{m}} \; + \; c \; \left| x\right|^2 \; \ln \left| x\right|
^2\quad ,\quad \lim_{x\rightarrow \infty
}\phi_{m,\widetilde{m}}(x\,,x^{*})\simeq
1+ \; c \; x^{m-1} \; x^{*\widehat{m}-1} \; \ln \left| x\right| ^{-2}
\]
and
\[
\lim_{x\rightarrow 1}\phi _{m,\widetilde{m}}(x\,,x^{*})\simeq
(1-x)^m \; (1-x^{*})^{\widehat{m}}+c\left| 1-x\right| ^2 \; \ln \left|
1-x\right|^2\,.
\]
For the function $\Psi _{m,\widetilde{m}}(\overrightarrow{p_1},\,%
\overrightarrow{p_2},\,\overrightarrow{p_3})$ in the limit $\left|
p_1\right| \rightarrow 0$ the region of large $\left| \rho
_{10}\right| $ is
essential. Taking into account only the singular terms in this limit, we
obtain
\[
\Psi _{m,\widetilde{m}}(\overrightarrow{p_1},\,\overrightarrow{p_2},\,%
\overrightarrow{p_3})\simeq \Psi _{m,\widetilde{m}}(\overrightarrow{p_2},\,%
\overrightarrow{p_3})\,\frac 12\,\left| p_1\right| ^2\ln \frac{\left|
P\right| ^2}{\left| p_1\right| ^2}
\]
where
\[
\Psi _{m,\widetilde{m}}(\overrightarrow{p_2},\,\overrightarrow{p_3})=
\]
\[
\prod_{r=2}^3\left| p_r\right| ^2\int \prod_{k=2}^3\left[
\frac{d^2\rho_k}{2\pi } \; \exp(i\overrightarrow{p_k} \cdot
\overrightarrow{\rho _{k0}}) \right]\;
\left( \frac{\rho_{23}}{\rho_{20}\rho_{30}}\right)^{m-1}
\left(\frac{\rho_{23}^{*}}{\rho_{20}^{*}\rho
_{30}^{*}}\right)^{\widetilde{m}-1}\partial \partial
^{*}\phi_{m,\widetilde{m}}(x\,,x^{*})
\]
and
\[
x=\frac{\rho _{30}}{\rho _{32}} \; .
\]
The last function can be simplified in the limit $p_2\rightarrow 0$,
corresponding to $\rho _{20}\rightarrow \infty $ and $x\rightarrow 0$.
Indeed, we can use the expansion
\[
\partial \partial ^{*}\phi _{m,\widetilde{m}}(x\,,x^{*})\simeq
m\widetilde{m}%
\,\,x^{m-1}(x^*)^{\widehat{m}-1}+c \; \ln \left| x\right| ^2+\ldots
\]
and verify, that the dependence from $\rho _{30}$ is canceled
in the contribution to the integrand from the first term
in the right hand side, leading to a vanishing
result after integration. Therefore taking into account only the
second term, we obtain,
\[
\Psi _{m,\widetilde{m}}(\overrightarrow{p_2},\,\overrightarrow{p_3})\sim
c_3\,.
\]
Hence, the resulting  behavior for the Odderon wave function at
$|p_1|<<|p_2|<<|p_3|$ is
\begin{equation}
\Psi_{m,\widetilde{m}}(\overrightarrow{p_1},\,\overrightarrow{p_2},\,
\overrightarrow{p_3})\sim c_3 \, \left| p_1\right| ^2 \; \ln \frac{\left|
P\right| ^2}{\left| p_1\right| ^2} \; .
\end{equation}
This is in agreement with the fact that it is an eigenfunction of the
integrals of
motion $q_2$ and $q_3$ provided that we take into account
in $\Psi _{m,\widetilde{m}}$ also the regular terms
proportional to $p_1$ and $p_1^*$.

It is natural to expect a similar behavior
\begin{equation}
\Psi _{m,\widetilde{m}}(\overrightarrow{p_1},\overrightarrow{p_2},
\ldots\,,
\overrightarrow{p_n})\sim c_n \, \left| p_1\right| ^2\ln \frac{\left|
P\right| ^2}{\left| p_1\right| ^2}
\end{equation}
for the case of $n$ reggeized gluons in the limit
$|p_1|<<|p_2|<<\ldots<<|p_n|$. Indeed, there are two independent
solutions for the eigenfunctions of the integrals of motion
$Q_j$ (\ref{intmov}) which behave at small $p_1$ correspondingly as
$f+g\, p_1\, \ln p_1$ and $g \, p_1$, where $f$ and $g$ are some
functions of $p_k$ analytic near $p_1=0$.  For the single-valuednes
property we should multiply two such functions depending on the
holomorphic
and anti-holomorphic variables. Further, because the operators $Q_j$
have more derivatives over $p_k$ ($k=2,3,...,n$) than the momenta
compensating them, $|g|^2$ should be a constant for small values of
these momenta.

In the opposite limit
\[
|p_n|<<|p_{n-1}|<<...<<|p_1|
\]
we obtain correspondingly
\begin{equation}
\Psi
_{m,\widetilde{m}}(\overrightarrow{p_1},\overrightarrow{p_2},\ldots \,
, \overrightarrow{p_n})\sim c_n \, \left| p_n\right| ^2\ln
\frac{\left|
P\right| ^2}{\left| p_n\right| ^2} \,.
\end{equation}

The fact that the behavior of $\Psi _{m,\widetilde{m}}$ at $\left|
p_n\right| \rightarrow 0$ for the composite state of $n$ reggeized
gluons is the  same as in the Pomeron case implies the existence of
a pole in
$$
\Psi_{m,\widetilde{m}}(\overrightarrow{\lambda _1},\,
\overrightarrow{\lambda
_2},\ldots,\overrightarrow{\lambda}_{n-1})
$$
at $\lambda_{n-1}=i $ and at $\lambda _{n-1}^{*}=i$.

Indeed, for $1=|p_1|>>|p_2|>>\ldots>>|p_n|$ we have
\bea
&&\Psi_{m,\widetilde{m}}(\overrightarrow{p_1},\overrightarrow{p_2},
\ldots\,,
\overrightarrow{p_n})= \\ \cr
&&2^{\frac{n-1}{2}}  \prod _{k=1}^{n-1}
\left( \int_{-\infty}^{+\infty} d \sigma _k
\sum _{N_k=-\infty}^{+\infty}  \,
\exp [i(t_k \lambda_k^*+t_k^*\lambda_k)]\right)\Psi _{m,\widetilde{m}}(
\overrightarrow{\lambda _1},\,\overrightarrow{\lambda
_2},\ldots,\overrightarrow{\lambda}_{n-1})  \nonumber
\eea
and therefore the contour of the integration in $\sigma _{n-1}$ for
$N_{n-1}=0$ should
be shifted in the complex plane up to the pole $(\sigma _{n-1}
-i)^{-2}$.

We can find the singular part of the Hamiltonian in the BS
representation for small $\left| p_n\right| $  similarly to the
Pomeron case:
\bea
&&\frac 12\left( H_{n,n-1}+H_{1n}\right) p_n^{-i\lambda
_{n-1}^*}p_n^{*-i\lambda
_{n-1}}=\frac 12\ln \left| p_1p_{n-1}\right| ^2p_n^{-i\lambda
_{n-1}^{*}}p_n^{*-i\lambda_{n-1}}+ \\ \cr
&&\left[ -\ln \left| p_n\right| ^2+\psi (1+i\lambda _{n-1}^{*})+\psi
(1-i\lambda _{n-1}
^{*})+\psi (1+i\lambda _{n-1})+\psi (1-i\lambda _{n-1})-4\psi
(1)\right]
\, p_n^{i\lambda _{n-1}^{*}}p_n^{*-i\lambda_{n-1}} \; . \nonumber
\eea
The first term in the right hand side can be combined with the other pair
hamiltonians. After that, we obtain for  $ 1
=|p_1|>>|p_2|>>\ldots>>|p_n|$:
\[
\left( \frac 12\ln \left| p_1p_{n-1}\right| ^2+\frac 12 \sum
_{r=1}^{n-2}
H_{r,r+1}\right)
\,\prod _{k=2}^{n-1}
p_k^{-i\lambda _k^*}p_k^{* -i\lambda _k}=
\]
\[
\frac{1}{2}\sum _{r=2}^{n-1}
\left[ \psi (1+i\lambda_r^{*})+\psi (1-i\lambda_r^{*})+\psi
(1+i\lambda_r)+\psi(1-i\lambda_r)-4\psi (1)\right] \,\prod_{k=2}^{n-1}
p_k^{-i\lambda_k^*}p_k^{*-i\lambda _k} \; .
\]
Thus, for the constant behavior
$\Psi_{m,\widetilde{m}}(\overrightarrow{p_1},\ldots,
\overrightarrow{p_{n-1}})\sim c$ at $ |p_2|>>\ldots>>|p_n| $
(corresponding to $\lambda _1=\lambda _1^{*}=\ldots= \lambda_{n-2}=
\lambda_{n-2}^*=0$) the last contribution vanishes and therefore we obtain
for the composite state energy the result similar to  the pomeron case
\be \label{nreggene}
E=i\lim_{\lambda ,\lambda ^{*}\rightarrow i}\left\{ \frac \partial
{\partial \lambda }\ln \left[ (\lambda -i)\,\lambda \,\Psi (\lambda
;\,m,\,\vec{\mu})\right]
+\frac \partial {\partial \lambda ^{*}}\ln \left[ (\lambda ^{*}-i)\,
\lambda
^* \; \Psi  (\lambda ^*;\,\widetilde{m}, \, \vec{\mu ^s})\right]
\right\} \; .
\ee
Here $\Psi (\lambda _{n-1};\,m,\,\vec{\mu})$ and $\Psi  (\lambda
^*_{n-1};\,,\widetilde{m}, \, \vec{\mu ^s})$ are
correspondingly holomorphic and anti-holomorphic factors of the wave
function at $\lambda _k=\lambda ^*_k=0, \; 1 \leq k \leq n-2 $:
$$
\Psi_{m,\widetilde{m}}(0,0,...,\overrightarrow{\lambda_{n-1}})
\Longrightarrow
\Psi (\lambda _{n-1};\,m,\,\vec{\mu})\,\Psi  (\lambda
^*_{n-1};\,\widetilde{m}, \, \vec{\mu ^s})
$$
and $\mu _k=(-i)^k q_k,\,\mu ^s_k=i^k q_k $ are eigenvalues of the
integrals of motion.
This quantity can be related with the Baxter function and the
normalization factor
for the pseudovacuum state [see eq. (\ref{pseudo})]:
$$
\Psi _{m,\widetilde{m}}(0,0,...,\overrightarrow{\lambda
_{n-1}}) \Longrightarrow c_{0,0,...,\overrightarrow{\lambda _{n-1}}}
^{ps}\,\; |\lambda
_{n-1}|^{2(n-1)} \; Q (\lambda _{n-1};\,m,\,\vec{\mu})\,Q  (\lambda
^*_{n-1};\,\widetilde{m}, \, \vec{\mu ^s})
\; .
$$
As it was argued above,
for the pseudovacuum state it looks plausible, that the correct
normalization of the kernel for the transition between momentum and
BS representations corresponds to
$c^{ps}_{0,0,...,\overrightarrow{\lambda _{n-1}}}=\sinh
^{n-2}(2\pi \,\lambda _{n-1})\,\sinh
^{n-2}(2\pi \,\lambda ^*_{n-1})$ [see eq.(\ref{psnorm})].
We obtain in this case for the energy
\be \label{engen}
E=i\lim_{\lambda ,\lambda ^{*}\rightarrow i}\left\{ \frac \partial
{\partial \lambda }\ln \left[ \sinh ^{n-1}(2\pi \,\lambda )\,\lambda
^n\,Q (\lambda ;\,m,\,\vec{\mu})
\right]
+\frac \partial {\partial \lambda ^{*}}\ln \left[ \sinh
^{n-1}(2\pi \,\lambda ^*)
\,\lambda
^{*n} \; Q  (\lambda
^*;\,\widetilde{m}, \, \vec{\mu ^s})\right] \right\} .
\ee
Thus, the energy is expressed in terms of
the behavior of the Baxter function $  Q(\lambda ,m) $ near $ \lambda
= i $.

\bigskip

In the customary case of spin chains the Baxter function is a
polynomial of degree $L$,
$$
Q_{XXX}(\lambda) = \prod_{k=1}^L (\lambda -\lambda_k)
$$
where the $ \lambda_k $ are solutions of the Bethe Ansatz
equations\cite{QISM,rev}. The energy of the XXX chain is given by,
\be \label{enXXX}
\left. E_{XXX}= -2 \sum_{k=1}^L { 1 \over \lambda_k^2 +1} = i {d \over d
\lambda } \log{ Q_{XXX}(\lambda+i) \over Q_{XXX}(\lambda-i) }
\right|_{\lambda=0}
\ee
In the present case the Baxter function is a meromorphic function with
an infinite number of poles and zeroes as discussed in sec. 5.2. It
can be expressed as an infinite product of the Bethe Ansatz
solutions [see eq.(\ref{repQpi})]. The eigenvalue expression here is
not given by eq.(\ref{enXXX}) but by eq.(\ref{enpom})
$$
E_{12} = 1 + \gamma + \psi(2-m) - \sum_{k=1}^\infty{\frac{1 }{i \, \lambda_k
\left( 1 + i \, \lambda_k\right)}} +  [m \to \widetilde{m}] \; ,
$$
in the pomeron case.

\section{BS representation for the Odderon wave function}

Here we consider the $\lambda$-representation for the Odderon, where
the operator
\[
B^{(3)}(u)=-P\left[ u^2+iu\left( \frac \partial {\partial t_1}+\frac
\partial {\partial t_2}\right) -\left( 1-e^{t_1-t_2}\right) \frac \partial
{\partial t_1}\frac \partial {\partial t_2}\right]
\]
is diagonal. Introducing the new variables,
\[
t=t_1+t_2=\ln \left[\frac{p_1\,(p_1+p_2)}{(p_2+p_3)\,p_3}\right] \quad ,
\quad z=e^y=e^{t_1-t_2}= \frac{p_1\,p_3}{(p_2+p_3)(p_1+p_2)}\; ,
\]
we obtain,
\[
B^{(3)}(u)=-P\left[ u^2+2iu\frac \partial {\partial t}+\left( 1-z\right)
\,z\frac \partial {\partial z}\,z\frac \partial {\partial z}-(1-z)\,\left(
\frac \partial {\partial t}\right) ^2\right] \; .
\]
To diagonalize $B^{(3)}$ one should find the eigenvalues and
eigenfunctions of the differential operators

\[
i\frac \partial {\partial t}\varphi =-\frac{\lambda _1^*+\lambda
_2^*}{2} \; \varphi \,,\,\,\,\left[
\left( 1-z\right) \,z\frac \partial {\partial z}\,
z\frac \partial {\partial
z}-z\,\left(\frac{\lambda _1^*+\lambda
_2^*}{2} \right)^2\right]
\varphi =-\left(\frac{\lambda _1^*-\lambda
_2^*}{2} \right)^2
\varphi \,,
\]
where $\lambda _k$ are the eigenvalues of the zeroes of
$B^{(3)}(u)$:

\[
B^{(3)}(u)=-P\prod_{k=1}^2\left( u-\widehat{\lambda }_k\right)
\,.
\]
The solution of the above equations can be written in terms of
hypergeometric functions
\[
\varphi _{\lambda _1^*\lambda
_2^*}(t\,,\,z)=e^{i\,\frac{\lambda _1^*+\lambda
_2^*}{2}
\,t}\,z^{i\,\frac{\lambda _1^*-\lambda
_2^*}{2} }F(-i\lambda _2^* \,,\,i\lambda _1^*
\,;\,1+i(\lambda _1^*-\lambda
_2^*) \,;\,z)=
\]
\[
e^{i\, \frac{\lambda _1^*+\lambda
_2^*}{2}  \,t}\,\,\frac{\Gamma
(1+i(\lambda _1^*-\lambda
_2^*)  )\;
z^{i\frac{\lambda _1^*-\lambda
_2^*}{2} }}{\Gamma (-i\lambda _2^* +1)\,\Gamma (i\lambda _1^*
)}\,\int_1^\infty \left(
\frac{x-1}{
x-z}\right) ^{-i \lambda _2^*  }\,x^{-i\lambda _1^*
-1}\,d\,x\,.
\]
An independent solution follows by interchanging here $ \lambda
_1^*$ and
$\lambda _2^* $.

Therefore, we can write the following relation between the wave
functions in
momentum and BS representations,
\[
\Psi_{m,\widetilde{m}}(\overrightarrow{p_1},\,\overrightarrow{p_2},\,
\overrightarrow{p_3})=
\]
\[
P^{\widetilde{m}}(P^*)^m \,\prod_{k=1}^2\left(\sum_{N_k=-
\infty }^{+\infty} \int_{-\infty }^{+\infty}
d\sigma_k\right)\,U_{\overrightarrow{\lambda
_1},\overrightarrow{\lambda _2}}(\overrightarrow{t},
\overrightarrow{z}) \; \Psi_{m,\widetilde{m}}
(\overrightarrow{\lambda_1},\,\overrightarrow{\lambda _2})\,,
\]
where
\be \label{alfbet}
\lambda_k=\sigma _k+i \, \frac{N_k}2 \; , \;
\lambda^*_k=\sigma _k-i \, \frac{N_k}2 \; .
\ee
and
\[
U_{\overrightarrow{\lambda_1},\overrightarrow{\lambda_2}}
(\overrightarrow{t},\overrightarrow{z})=C_{\overrightarrow{\lambda_1
},\overrightarrow{\lambda_2}} \;\;
e^{it \frac{\lambda _1^*+\lambda
_2^*}{2}  } \; e^{it^{*}\frac{\lambda _1+\lambda
_2}{2} } \; \;
U _{\overrightarrow{\lambda_1},
\overrightarrow{\lambda_2}}(\overrightarrow{z})\; .
\]
Here the function
$U_{\overrightarrow{\lambda_1},\overrightarrow{\lambda_2}}(
\overrightarrow{z} ) $ is defined as follows
\be \label{defiPhi}
U_{\overrightarrow{\lambda_1},\overrightarrow{\lambda_2}}(\overrightarrow{z}
)\,=z^{i\frac{\lambda _1^*-\lambda
_2^*}{2}  }\; (z^*)^{i\frac{\lambda _1-\lambda
_2}{2}  }\int
\frac{d^2x}{\left| x\right| ^2}\,
\,x^{-i\lambda _1^* } \; (x^*)^{-i\lambda _1}\,\left( \frac{x-1%
}{x-z} \right) ^{-i\lambda _2^* }\left(
\frac{x^{*}-1}{x^{*}-z^{*}}\right)^{-i\lambda _2}
\ee
and the normalization constant
$C_{\overrightarrow{\lambda_1},\overrightarrow{\lambda_2
}}$ can be found from the orthogonality condition
\be \label{conorodd}
\int {d^2t \,d^2z \over |z(1-z)|^2 } \; U
_{\overrightarrow{\lambda_1 },\overrightarrow{\lambda_2}}(
\overrightarrow{t},\overrightarrow{z}) \;
U_{\overrightarrow{\lambda_1
^{\prime }},\overrightarrow{\lambda_2^{\prime }}}^{*}(\overrightarrow{t},
\overrightarrow{z})=\sum _P\prod_{k=1}^2\delta (\sigma _k-\sigma
_{i_k}^{\prime})\,\delta _{N_k,N_{i_k}^{\prime }}.
\ee
It should be taken into account, that due to the symmetry properties
of $U_{\overrightarrow{\lambda_1 },\overrightarrow{\lambda_2}}(
\overrightarrow{z} ) $ under
$\lambda_1 \leftrightarrow \lambda_2$  two terms appear in the right
hand side of the orthogonality equation.

Let us show that the kernel of the unitary transformation $
U_{\overrightarrow{\lambda_1
},\overrightarrow{\lambda_2 }}(\overrightarrow{t}, \overrightarrow{z}) $
has an interpretation in terms of the Feynman diagram as it was
in the case of  the Pomeron wave function (\ref{pomdiag}). After changing the
integration variable $ x $ into $ k $ as follows,
$$
x = {p_1 \over 1-p_1} \; {k \over  1-k}
$$
eq.(\ref{defiPhi}) takes the form
\bea
&&{U_{\overrightarrow{\lambda_1},\overrightarrow{\lambda_2
}}(\overrightarrow{t}, \overrightarrow{z}) \over
C_{\overrightarrow{\lambda_1 },\overrightarrow{\lambda_2  } }}
= \left({p_1 \over p_3}\right)^{i
\lambda^*_2}\;  \left({p_1^* \over p_3^*}\right)^{i\lambda_2}
\int {d^2k \over |k(1-k)|^2 }  \cr \cr
&& \left( {k \over 1-k}\right)^{-i\lambda^*_1}
\; \left( {k^* \over 1-k^*}\right)^{-i\lambda_1} \;
 \left( {k+p_1-1 \over k - p_3}\right)^{-i\lambda^*_2}\;
 \left( {k^*+p^*_1-1 \over k^* - p^*_3}\right)^{-i\lambda_2}.
\eea
The wave function in the $\lambda$-representation has the form
$$
 \Psi_{m,\widetilde{m}}
(\overrightarrow{\lambda_1},\,\overrightarrow{\lambda
_2})=
$$
$$
\int
\frac{d^2p_1d^2p_3}{|p_1|^2|1-p_1-p_3|^2|p_3|^2}\,
U^*_{\overrightarrow{\lambda
_1},\overrightarrow{\lambda
_2}}(\overrightarrow{t},
\overrightarrow{z}) \;\Psi_{m,\widetilde{m}}
(\overrightarrow{p_1},\,\overrightarrow{1}-\overrightarrow{p_1}-
\overrightarrow{p_3},\,
\overrightarrow{p_3})
$$
The associated Feynman diagram is depicted in fig. 1.

\begin{figure}[h]
\epsfig{file=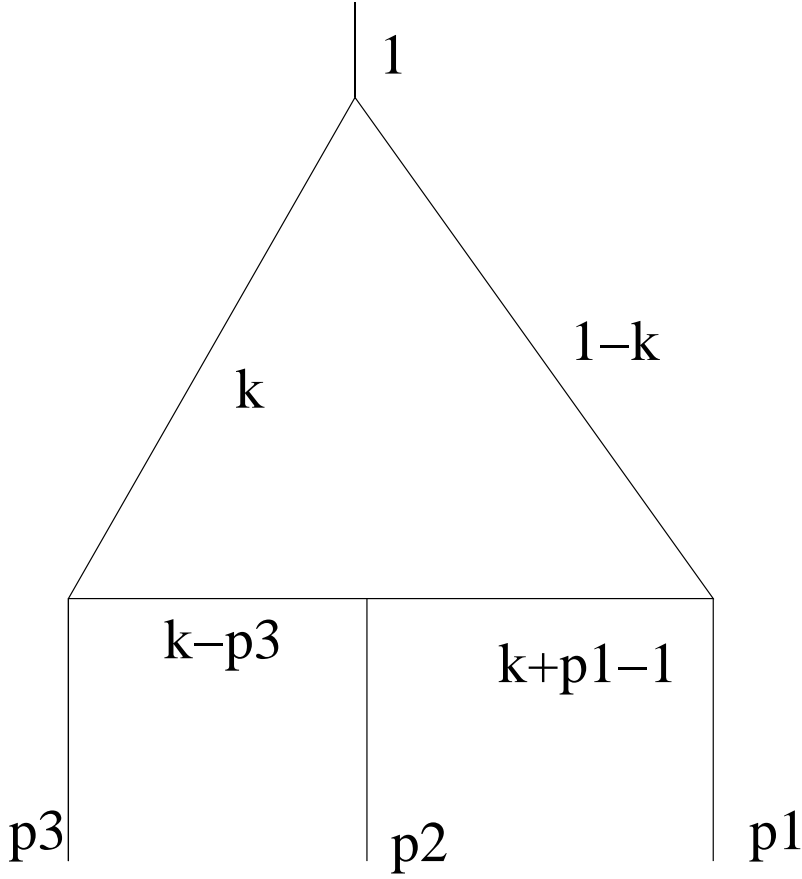}
\caption{Odderon Feynman diagram}
\end{figure}

\section{Properties of the unitary transformation for the Odderon
wave function}

The function $U_{\overrightarrow{\lambda _1
},\overrightarrow{\lambda _2}}(
\overrightarrow{z} ) $ is a solution of the differential equations of
a hypergeometric type in both variables $ z $ and $ z^* $,
\bea
&&\left[ \left( 1-z\right) \,z\frac d {d z}\,z\frac d
{dz}-z\,\left(\frac{\lambda _1^*+\lambda
_2^*}{2}\right)^2 + \left(\frac{\lambda _1^*-\lambda
_2^*}{2}\right)^2
\right]U = 0 \; , \cr \cr
&&\left[ \left( 1-z^*\right) \,z^*\frac d {d z^*}\,z^*\frac d
{dz^*}-z^*\, \left(\frac{\lambda _1+\lambda
_2}{2}\right)^2 +
\left(\frac{\lambda _1-\lambda
_2}{2}\right)^2
\right] U=0  \,, \nonumber
\eea
Therefore,
$U_{\overrightarrow{\lambda
_1},\overrightarrow{\lambda _2}}(\overrightarrow{z}
) $ is a bilinear
combination of independent solutions being functions of $z$ and  $
z^* $.
In addition,
$\Phi_{\overrightarrow{\alpha},\overrightarrow{\beta}}(\overrightarrow{z})
$ should be a single valued function of $\overrightarrow{z}$ near
the
singularities $ |z| = 0 \; ,  |z-1| = 0\; , |z| =\infty $. The
effective way to satisfy these requirements  \cite{DF} is to use the
monodromy
properties \cite{gr} of the two independent solutions:
$$
z^{i\frac{\lambda _1-\lambda
_2}{2}}F(-i\lambda _2 \,,\,i\lambda _1 \,;\,1+i(\lambda
_1-\lambda
_2) \,;\,z) \,\,,\,\,\,  z^{-i\frac{\lambda _1-\lambda
_2}{2}}F(-i\lambda_1 \,,\,i\lambda _2 \,;\,1-i(\lambda _1-\lambda
_2) \,;\,z)
$$
and analogous expressions in $ z^* $. Thus, we obtain,
\be\label{Phiexpl}
U_{\overrightarrow{\lambda _1 },\overrightarrow{\lambda _2}}(
\overrightarrow{z} ) = K_{\overrightarrow{\lambda _1
},\overrightarrow{\lambda _2}} \left[ \chi _{\lambda _1 \lambda _2}
(z^*)\chi  _{\lambda _1^*\lambda _2^*}(z)-\chi  _{\lambda
_2 \lambda _1}(z^*)\chi
 _{\lambda _2^* \lambda _1^*}(z) \right] \; ,
\ee
where
\bea \label{defxi}
\chi  _{\lambda _1 \lambda _2}(z^*) &\equiv& a _{\lambda _1 \lambda
_2}
\,\,(z^*)^{i\frac{\lambda _1-\lambda
_2}{2} }\,F(-i\lambda _2
\,,\,i\lambda _1
\,;\,1+i(\lambda _1-\lambda _2) \,;\,z^*) \quad , \cr \cr
a_{\lambda _1 \lambda _2 }&\equiv&\frac{\Gamma (i\lambda _1
)\Gamma
(-i\lambda _2 )}{\Gamma (1+i(\lambda _1-\lambda _2) )} \; ,
\eea
and an analogous expression for $ \chi _{\lambda _1^* ,\lambda _2^*
}(z) $. This result can be verified by the direct calculation of
the integral (\ref{defiPhi}). In such way we obtain for
the constant $ K_{\overrightarrow{\lambda
_1},\overrightarrow{\lambda _2}} $ in
eq.(\ref{Phiexpl}):

$$
 K_{\overrightarrow{\lambda _1},\overrightarrow{\lambda _2}} = i
\left| \lambda _2  \right|^2 \; { \sinh (\pi \lambda _2) \;
\sinh (\pi \lambda _1 ) \over \sinh ( \pi (\lambda _1-\lambda _2))
}\,,
$$
where we used [see eqs.(\ref{alfbet})]
\[
{ \sinh (\pi \lambda _2 ) \; \sinh (\pi \lambda _1 ) \over
\sinh ( \pi
(\lambda _1-\lambda _2))} ={ \sinh (\pi \lambda _2^* ) \; \sinh (\pi
\lambda _1 ^*) \over
\sinh ( \pi
(\lambda _1^*-\lambda _2^*))}  .
\]

In summary, collecting all factors we find for the matrix elements of
the unitary transformation relating momentum and BS representations,
$$
{U_{\overrightarrow{\lambda _1 },\overrightarrow{\lambda _2
}}(\overrightarrow{t},
\overrightarrow{z}) \over C_{\overrightarrow{\lambda _1
},\overrightarrow{\lambda _2
} } \;K_{\overrightarrow{\lambda _1},\overrightarrow{\lambda _2}}}=
$$
\be \label{BSodd}
e^{i( t\frac{\lambda _1^*+\lambda
_2^*}{2}  +
t^{*}\frac{\lambda _1+\lambda
_2}{2})} \; \left[ \chi _{\lambda _1 \lambda _2}
(z^*)\chi  _{\lambda _1^*\lambda _2^*}(z)-\chi  _{\lambda
_2 \lambda _1}(z^*)\chi
 _{\lambda _2^* \lambda _1^*}(z) \right]\; ,
\ee
where $\chi _{\lambda _1 \lambda _2}
(z^*)$ and $\chi  _{\lambda _1^*\lambda _2^*}(z)$ are given
by
eq.(\ref{defxi}).

We find for $ z \to 0 $ the asymptotic behavior,
$$
\frac{U_{\overrightarrow{\lambda _1 },\overrightarrow{\lambda _2
}}(\overrightarrow{t},
\overrightarrow{z})}{C_{\overrightarrow{\lambda _1
},\overrightarrow{\lambda _2
} } \,K_{\overrightarrow{\lambda _1},\overrightarrow{\lambda _2}}}=
$$
$$
e^{i( t\frac{\lambda _1^*+\lambda
_2^*}{2}  +
t^{*}\frac{\lambda _1+\lambda
_2}{2})} \,\left[ a _{\lambda _1 \lambda _2}
a _{\lambda _1^*\lambda _2^*}\,z^{*i\frac{\lambda
_1-\lambda
_2}{2}} \,z^{i\frac{\lambda _1^*-\lambda
_2^*}{2}}-a
_{\lambda
_2 \lambda _1} a
 _{\lambda _2^* \lambda _1^*}\, z^{*-i\frac{\lambda _1-\lambda
_2}{2}} z^{-i\frac{\lambda _1^*-\lambda
_2^*}{2}}\right]\,.
$$

\bigskip

While the phases of the constant factors in eq.(\ref{BSodd}) are
special functions, the squared modulus is quite simple. Indeed, we find

$$
\left| a_{\lambda _1 ,\lambda _2 } \; a_{\lambda _1^* ,\lambda _2^*
} \;
  K_{\overrightarrow{\lambda _1},\overrightarrow{\lambda_2}} \right|
= { \pi
  \over \sqrt2} \left| { \lambda_2 \over\lambda_1(\lambda_1 -
  \lambda_2) }\right| \; .
$$
The normalization condition (\ref{conorodd}) then yields,
\be \label{codd}
C_{\overrightarrow{\lambda _1 },\overrightarrow{\lambda _2 } }\sim
\left| {
  \lambda_1 \over\lambda_2}(\lambda_1 - \lambda_2) \right|
\ee
up to a numerical constant.

\bigskip

Using the relation between hypergeometric functions with
mutually inversed arguments \cite{gr}  we obtain for
$ z\rightarrow \infty $
\bea
&&\chi_{\lambda _1 ,\lambda _2 }(z^*)\buildrel{ |z| \to
\infty}\over={\Gamma
(i(\lambda _1+\lambda _2) )\,\Gamma (-i\lambda _2 ) \over \Gamma
(1+i\lambda _1)}
\,(z^*)^{i\frac{\lambda _1+\lambda _2}{2} } \;
e^{i\pi \lambda _2\, \frac{Im(z)}{|Im(z)|}}+
\cr \cr
&&+ {\Gamma (-i(\lambda _1+\lambda _2) )\Gamma(i\lambda _1 )
\over \Gamma (1-i\lambda _2 )} \,(z^*)^{-i\frac{\lambda
_1+\lambda _2}{2}} \; e^{-i\pi \lambda _1\,
\frac{Im(z)}{|Im(z)|}} \; . \nonumber
\eea
As a consequence of this asymptotic behavior, the interference terms
in $\Phi_{ \overrightarrow{\lambda _1 },
\overrightarrow{\lambda _2}} (\overrightarrow{z}) $ cancel using the
relation
\[
\sinh (\pi \lambda _1) \sinh (\pi \lambda _2^*)=\sinh (\pi
\lambda _2)\sinh (\pi \lambda _1 ^{*}) \; .
\]
We obtain with the use of the equality
\[
\sinh (\pi \lambda _1 )\sinh (\pi \lambda _1^*)-\sinh
(\pi
\lambda _2)\sinh (\pi \lambda _2^{*})=
\]
\[
e^{\pi (\lambda _2-\lambda _2^*)}\sinh (\pi (\lambda _1-\lambda _2)
)\sinh
(\pi (\lambda _1+\lambda _2) )
\]
the following asymptotics for $U _{\overrightarrow{\lambda _1 },
\overrightarrow{\lambda _2 }}(\overrightarrow{z})$ for large $|z|$
\[
U _{\overrightarrow{\lambda _1 },\overrightarrow{\lambda _2
}}(\overrightarrow{z}
)\,=b_{\overrightarrow{\lambda _1 },\overrightarrow{\lambda _2
}}^{(1)} \;
(z^*)^{i\frac{\lambda _1+\lambda _2}{2}}\,z^{i\frac{\lambda
_1^*+\lambda _2^*}{2}} +b_{\overrightarrow{\lambda _1
},\overrightarrow{\lambda _2
}}^{(2)} \;
(z^*)^{-i\frac{\lambda _1+\lambda _2}{2}}\,z^{-i\frac{\lambda
_1^*+\lambda _2^*}{2}}
\]
where
\[
b_{\overrightarrow{\lambda _1 },\overrightarrow{\lambda _2
}}^{(1)}=\frac{\pi
\,\Gamma (-i\lambda _1)\Gamma (1-i\lambda _2 )\Gamma
(i(\lambda _1^{*}+\lambda _2^*))}{\Gamma (1-i(\lambda _1+\lambda _2
))\Gamma (i\lambda _2^{*})
\Gamma(1+i\lambda _1 ^{*})}\,,
\]
\[
b_{\overrightarrow{\lambda _1 },\overrightarrow{\lambda _2
}}^{(2)}=-\frac{\pi
\,\Gamma (i\lambda _2^*)\Gamma (1+i\lambda _1^* )\Gamma
(-i(\lambda _1+\lambda _2))}{\Gamma (1+i(\lambda _1^*+\lambda _2^*
))\Gamma (-i\lambda _1)
\Gamma(1-i\lambda _2 )}\,.
\]
This can also be obtained from the integral (\ref{defiPhi}) by direct
calculation.

We analogously find using the series expansion for the hypergeometric
function when $ c = a + b + 1 $ \cite{abra}
\bea
&&\chi_{\lambda _1,\lambda _2 }(z^*) =
(z^*)^{i\frac{\lambda _1-\lambda _2}{2}}\left\{ {1 \over \lambda _1
\lambda _2} +{ 1-z^* \over \Gamma (1-i\lambda _2 )\Gamma
(1+i\lambda _1 )} \right. \times \cr \cr
&& \sum_{n=0}^{\infty} { \Gamma (n+1-i\lambda _2 )\Gamma
(n+1+i\lambda _1 ) \over n! (n+1)!} (1-z^*)^n  \times \cr \cr
&&\left. \left[ \log(1-z^*) - \psi(n+1)
 -\psi(n+2)  + \psi(n+1-i\lambda _2 )  + \psi(n+1+i\lambda _1)
\right] \right\} \; . \nonumber
\eea
We obtain in the limit $z^*\rightarrow 1$:
\bea
&&\lim_{z^*\rightarrow 1}\chi _{\lambda _1 ,\lambda _2 }(z^*)=\frac
1{\lambda _1 \lambda _2} + \cr \cr
&&+(1-z^*)\left[ \ln (1-z^*)+\psi (1-i\lambda _2 )+\psi (1+i\lambda
_1
)-\psi (1)-\psi (2) + {i \over 2\lambda _1}- {i  \over 2\lambda _2
} \right] \nonumber
\eea
in agreement with eq.(\ref{tychic}). Thus we obtain,
\begin{equation}
\lim_{z\rightarrow 1}
{U_{\overrightarrow{\lambda _1 },\overrightarrow{\lambda _2
}}(\overrightarrow{z})
\over K_{\overrightarrow{\lambda _1},\overrightarrow{\lambda _2}}} =
-i\pi \frac{\sinh (\pi (\lambda _1 -\lambda _2))}{\sinh (\pi
\lambda _1) \sinh (\pi \lambda _2)} \left( \frac{1-z}{\lambda _1
\lambda
_2}+\frac{1-z^{*}}{\lambda
_1^* \lambda _2^*
}+ \left|1-z\right| ^2 \ln \left| 1-z\right| ^2\right) \,,
\label{fiz1}
\end{equation}
where we used the relations
$$
 \psi (1-i\lambda _2 )+\psi (1+i\lambda _1 )-\psi
(1-i\lambda _1 )-\psi (1+i\lambda _2 )+\frac{i}{\lambda
_1}-\frac{i}{\lambda _2}=-i\pi \frac{\sinh (\pi (\lambda _1 -\lambda
_2))}{\sinh (\pi
\lambda _1) \sinh (\pi \lambda _2)}\,.
$$
Again,  the result (\ref{fiz1}) can be obtained directly from the integral
representation (\ref{defiPhi})
for $\Phi _{\overrightarrow{\alpha },\overrightarrow{\beta }}(
\overrightarrow{z})$. Taking into account the found above value of
the normalization constant we have for large momenta $p_k$ and fixed
$P$
\begin{equation} \label{oddass}
\frac{U_{\overrightarrow{\lambda _1 },\overrightarrow{\lambda _2
}}(\overrightarrow{t},\, \overrightarrow{z})}{\pi |\lambda _1
\lambda _2 (\lambda _1 -\lambda _2)|}=e^{i( t\frac{\lambda
_1^*+\lambda
_2^*}{2}  +
t^{*}\frac{\lambda _1+\lambda
_2}{2})} \,\left( \frac{1-z}{\lambda _1 \lambda
_2}+\frac{1-z^{*}}{\lambda
_1^* \lambda _2^*
}+ \left|1-z\right| ^2 \ln \left| 1-z\right| ^2\right)\,.
\end{equation}

\section{Acknowledgements}

We thank E. A. Antonov, A. P. Bukhvostov,  L. D. Faddeev and further
participants to the Winter School of the Petersburg Nuclear Physics
Institute (February 2001) for helpful discussions on the basic
results of this paper. Subsequent discussions with J. Bartels,
V. Fateev, R. A. Janik, G. Korchemsky, P. Mitter. A. Neveu, F. Smirnov,
J. Wosiek, A. Zamolodchikov and Al. Zamolodchikov were
especially fruitful. One of us (LNL) thanks LPTHE for the hospitality
during his visits to the University of Paris VI in January, February
and May 2001. When this paper was written, we learn
from G. Korchemsky that similar results were obtained by him in
collaboration with A. Derkachev and A. Manashov.

\section{Appendix A}

We compute in this Appendix the integral in eq.(\ref{fdopom}).

To start we need the Fourier transform\cite{gr},
\begin{equation}  \label{L2}
\int {\frac{d^2z }{2 \pi }} \; e^{i{\vec q}\cdot {\vec z}}\; z^{m} \;
(z^*)^{\widetilde m} = \frac{i^{{\widetilde m}-m} }{2^{-m-{\widetilde
m}-1}} \; q^{-{\widetilde m}-1} \; (q^*)^{-m-1}  \;
\frac{\Gamma(1+{\widetilde m})}{\Gamma(-m)}
\end{equation}

Entering the factor $ \left( \overrightarrow{p_1}\right)^2\left(
\overrightarrow{p_2}\right)^2 $ inside the integral in
eq.(\ref{fdopom}) as $ \nabla_1^2 \, \nabla_2^2 $ we find after
partial integration,
\begin{equation}  \label{inter}
\Psi_{m,\widetilde{m}}(\overrightarrow{p_1},\,\overrightarrow{p_2}
)= |m(m-1)|^2 \;
\int {\frac{d^2z_1 }{2 \pi}}
{\frac{d^2z_2 }{2 \pi}} \; e^{i\left({\vec p}_1\cdot {\vec z}_1 + {\vec p}
_2\cdot {\vec z}_2 \right)} \; \frac{(z_1-z_2)^{m-2}}{(z_1 \; z_2 )^m}
\frac{(z^*_1-z^*_2)^{{\widetilde m}-2}}{(z^*_1 \; z^*_2 )^{\widetilde m}}
\end{equation}
Now, we replace the $z$-factors in the integrand by the integral
representation (\ref{L2}):
$$
z^{-m} \; (z^*)^{-{\widetilde m}} = {i^{m-{\widetilde m}} \over
2^{m+{\widetilde m}-1}}  {\Gamma(1-m) \over \Gamma({\widetilde m})}
\int {\frac{d^2 k }{2 \pi }} \; e^{i{\vec q}\cdot {\vec k}}\; (k^*)^{m-1} \;
k^{{\widetilde m}-1} \; .
$$
The $z$-integrals in eq.(\ref{inter}) give now Dirac delta functions
and we obtain eq.(\ref{psipom}).

\section{Appendix B}

We derive here the asymptotic behavior of the Baxter function for the
Pomeron starting from the integral representation (\ref{funba}). We change
the integration variable to
\[
y \equiv 2 \mbox{ArgTanh}\left(1 - 2 p\right) \quad , \quad p = \frac12
\left( 1 - \tanh{\frac{y }{2}} \right)
\]
and obtain,
\[
Q(\lambda ,m) = i \; \frac{\pi \,\sinh (\pi \lambda )}{\sin (\pi \,m)}
\int_{-\infty}^{+\infty} dy \; e^{i \lambda \, y} \; P_{m-1}\left(\tanh{%
\frac{y }{2}}\right)
\]
The function $P_{m-1} (z) $ has a cut running from $z = - 1 $ till $z =
-\infty $. Therefore, $P_{m-1}\left(\tanh{\frac{y }{2}} \right) $ has cuts
in the $y$-plane from $y = i (2n+1)\, \pi $ till $y = -\infty $ where $n$ is
an integer. We now deform the integration path around the cut from $y = i\,
\pi $ till $y = -\infty $ and we find,
\begin{equation}  \label{defint}
Q(\lambda ,m) = i \; \frac{\pi \,\sinh (\pi \lambda )}{\sin (\pi \,m)}\;
e^{-\pi \, \lambda} \; \int_0^{+\infty} dx \; e^{i \lambda \, x} \left[
P_{m-1}\left(- \coth{\frac{x }{2}} + i 0 \right) - P_{m-1}\left(- \coth{%
\frac{x }{2}} - i 0 \right) \right]
\end{equation}
where we changed the integration variable as $y = i\, \pi - x $. The
integral (\ref{defint}) is dominated for large $\lambda $ by the end-point
$%
x = 0 $. Therefore, we insert in eq.(\ref{defint}) the representation of $%
P_{m-1} (z) $ appropriate for large $z = - \coth{\frac{x }{2}} \pm i 0
$\cite{gr}
\begin{eqnarray}
P_{m-1} (z) &=& {\frac{ \tan\pi m \; \Gamma(m) }{2^m \; \sqrt{\pi} \;
\Gamma\left(m + \frac12 \right) }}\; z^{-m-2}\;
_2F_1\left({\frac{m+1}{2}},{%
\frac{m }{2}}; m + \frac12 ; {\frac{1 }{z^2}}\right) +\cr \cr &+& {\frac{
2^{m-1} \, \Gamma\left(m - \frac12 \right) }{\sqrt{\pi} \; \Gamma\left(m
\right)}}\; z^{m-1} \; _2F_1\left({\frac{1-m}{2}},1-{\frac{m }{2}}; -m +
\frac32 ; {\frac{1 }{z^2}}\right)
\end{eqnarray}
Keeping here the dominant terms for large $z $ and using the relation
\[
\left(- \coth{\frac{x }{2}} + i 0\right)^{-m} - \left(- \coth{\frac{x }{2}}
- i 0\right)^{-m} = -2i\; \sin \pi m \left(\coth{\frac{x }{2}}\right)^{-m}
\quad ,\quad x>0 \quad ,
\]
we get for $\lambda \gg 1$ (for Re $m > 1/2 $)
\begin{eqnarray}  \label{asintQ}
Q(\lambda ,m)  &=& 4 \, \sqrt{\pi} \; \left(
4 \, i \; \lambda \right)^{m-2} \; {\frac{ \Gamma\left(m - \frac12 \right)
\; \Gamma\left(2-m \right) }{\Gamma\left(m \right) }} \left\{ 1 + {\cal O}
\left( {\frac{1 }{\lambda^2}}\right) + \right. \\ \cr
&+&\left. \left( 4 \,
i \; \lambda\right)^{1-2m} \; \tan \pi m \; {\frac{ \Gamma^2\left(m \right)
\; \Gamma\left(m+1 \right) }{\Gamma\left(m + \frac12 \right)\; \Gamma\left(m
- \frac12 \right)\; \Gamma\left(2-m \right) }} \left[ 1 + {\cal O}\left( {%
\frac{1 }{\lambda^2}}\right)\right] \right\} \; . \nonumber
\end{eqnarray}
For Re $m < 1/2 $ one should just replace $m \Rightarrow 1 - m $.

The case $m = 1/2 $ follows by taking the limit $m \to 1/2 $ in eq.(\ref
{asintQ}) with the result for $\lambda \gg 1$
\[
Q(\lambda ,\frac12 ) =
\frac{ \sqrt{\pi}}{%
2 \; \left( i \; \lambda \right)^{3/2}}\left\{ \left[ \log\left( {\frac{i
\, \lambda}{4}} \right) + 2 - 3 \gamma \right] + {\cal O} \left(
\frac{1 }{\lambda^2}\right) \right\} \, .
\]

\bigskip

Let us now derive the asymptotic behavior of $Q(\lambda ,m) $ starting from
their infinite product representation (\ref{repQpi}) and the asymptotic
distribution of their zeros (\ref{asiceros}).

For $\lambda \gg 1 $ the product will be dominated by zeros of the order $%
\sim \lambda $. We can then write,
\[
\prod_{k=1}^\infty \left(1 - {\frac{\lambda }{\lambda_k}}\right) e^{\frac{%
\lambda }{\lambda_k}} \simeq \prod_{k=1}^M \left[\left({\frac{1 - {\frac{%
\lambda }{\lambda_k}} }{1 +i {\frac{\lambda }{k+1-m}}}}\right) e^{{\frac{%
\lambda }{\lambda_k}} + {\frac{ i \lambda }{k+1-m}}}\right]
\prod_{k=1}^\infty \left(1 + {\frac{i \lambda }{k+1-m}}\right) e^{ - {\frac{
i \lambda }{k+1-m}}}
\]
where $M$ is a cutoff $1 \ll M \ll |\lambda| $. We obtain for $\lambda \gg 1
$ using the formulae \cite{gr}
\[
\prod_{k=1}^\infty \left(1+{\frac{iy }{k+x}}\right) e^{-{\frac{iy}{k}}}=
e^{-i \gamma \, y} {\frac{ \Gamma(1+x) }{\Gamma(1+x+iy) }}
\]
\[
\psi(x+1) + \gamma = \sum_{k=1}^{\infty} {\frac{x }{k(x+k)}} \; ,
\]
\begin{equation}
Q(\lambda ,m) = \mbox{constant} \;
\lambda^{m-2}
\end{equation}
in perfect agreement with eq.(\ref{asintQ}).

\end{document}